\documentclass{aa}
\usepackage[]{natbib}
\bibpunct{(}{)}{;}{a}{}{,} % to follow the A&A style
\usepackage{graphics}   % standard graphics specifications
\usepackage{graphicx}   % alternative graphics specificati.jpgons
\usepackage{epstopdf}
\usepackage{longtable}   % helps with long table options
\usepackage{url}      % for on-line citations
\usepackage{bm}        % special 'bold-math' package
\usepackage[varg]{txfonts}
\usepackage{pdflscape}
\usepackage{supertabular}
\usepackage{hyperref}   %link
\usepackage{wasysym}
\usepackage[svgnames]{xcolor}
\usepackage{amsmath}
\usepackage{amssymb}  %%% MAA loaded to have access to all math-symbols
\usepackage{nccmath}

\usepackage{circledsteps}

\hypersetup{
    unicode=false,          % non-Latin characters in Acrobat?s bookmarks
    colorlinks=true,       % false: boxed links; true: colored links
    linkcolor=blue,          % color of internal links (change box color with linkbordercolor)
    citecolor=blue,        % color of links to bibliography
    filecolor=blue,      % color of file links
    urlcolor=blue           % color of external links
}
\usepackage{color}

\usepackage[utf8]{inputenc}
\usepackage{lmodern}
\usepackage{tocloft}
\usepackage{enumitem}
\usepackage{enumerate}
\usepackage{graphicx}
\usepackage{appendix}
\usepackage{multirow}
\usepackage{array}
\usepackage{makecell}
\usepackage{cancel}
\usepackage{fancyhdr}
\usepackage{tikz}
\usetikzlibrary{decorations.markings,calc,patterns,decorations.pathmorphing}

\usepackage{float}
\usepackage{indentfirst}

\usetikzlibrary{shapes}
\usepackage{afterpage}

\usepackage{soul}

\newcounter{maacounter}
\newenvironment{maaenvironment}[1][]{\refstepcounter{maacounter} \nobreakspace 
   {\color{Sienna}{MAA(\themaacounter):~{#1}}} \rmfamily}{}
\makeatletter
\newcommand*{\rom}[1]{\expandafter\@slowromancap\romannumeral #1@}
\makeatother

\begin{document}

\title{Evolving massive stars to core collapse with GENEC: Extension of equation of state, opacities and effective nuclear network.}

\author{Adam Griffiths\inst{1}, 
Miguel-Á. Aloy\inst{1,2}, Raphael Hirschi\inst{3,4}, Moritz Reichert\inst{1}, Martin Obergaulinger\inst{1},
Emily E. Whitehead\inst{3},
Sebastien Martinet\inst{5},
Luca Sciarini\inst{6},
Sylvia Ekström\inst{6},
Georges Meynet\inst{6}
}
\titlerunning{Extension of Equation of state, opacities, and reduced nuclear network}
 \authorrunning{Griffiths et al.}

 \institute{Departament d'Astonomia i Astrof\'{\i}sica, Universitat de Val\`encia, C/Dr. Moliner, 50, E-46100 Burjassot (Val\`encia), Spain
 \\
 \email{adam.griffiths@uv.es}; \email{miguel.a.aloy@uv.es}
 \and
 Observatori Astronòmic, Universitat de València, 46980 Paterna, Spain
 \and
 Astrophysics Group, Lennard-Jones Laboratories, Keele University, Keele ST5 5BG, UK
 \and
 Kavli IPMU (WPI), University of Tokyo, 5-1-5 Kashiwanoha, Kashiwa 277-8583, Japan
 \and
 Institut d'Astronomie et d'Astrophysique, Universit\'e Libre de Bruxelles (ULB), CP 226, B-1050 Brussels, Belgium \and
Geneva Observatory, University of Geneva, Chemin Pegasi 51, CH-1290 Sauverny, Switzerland}

\date{Received /
Accepted}
\abstract
{Stars with initial mass above roughly 8 $M_{\odot}$ will evolve to form a core made of iron group elements, 
at which point no further exothermic nuclear reactions  between charged nuclei may prevent the core collapse. Electron capture, neutrino losses, and the photo-disintegration of heavy nuclei trigger the collapse of these stars.  
Models at the brink of core collapse are produced using stellar evolution codes, and these pre-collapse models may be used in the study of the subsequent dynamical evolution (including their explosion as supernovae and the formation of compact remnants such as neutron stars or black holes).} 
{We upgraded the physical ingredients employed by the GENeva stellar Evolution Code, GENEC, so that it covers the regime of high-temperatures and high-densities required to produce the progenitors of core-collapse. Our ultimate goal is producing pre-supernova models with GENEC, not only right before collapse, but also during the late phases (silicon and oxygen burning).}
{We have improved GENEC in three directions: equation of state, the nuclear reaction network, and the radiative and conductive opacities adapted for the computation of the advanced phases of evolution. We produce a small grid of pre-supernova models of stars with zero age main sequence masses of $15M_{\odot}$, $20M_{\odot}$, and $25M_{\odot}$ at solar and less than half solar metallicities. The results are compared with analogous models produced with the MESA code.}
{The global properties of our new models, particularly of their inner cores, are comparable to models computed with MESA and pre-existing progenitors in the literature. Between codes the exact shell structure varies, and impacts explosion predictions.}
{Using GENEC with state-of-the-art physics, we have produced massive stellar progenitors prior to collapse. These progenitors are suitable for follow-up studies, including the dynamical collapse and supernova phases. Larger grids of supernova progenitors are now feasible, with the potential for further dynamical evolution.}


\maketitle

\noindent
\section{Introduction}

Building stellar evolution models to reproduce observations remains a difficult task, due to a relatively large amount of free parameters and approximations used in the modelling.  An increase in the quality and quantity of observations of stellar objects, \citep[e.g. the increasing catalogue of observations by the GAIA collaboration,][]{Gaia_Collaboration_2016, Gaia_Collaboration_2018, Gaia_Collaboration_2021}, refines the global constraints that stellar models must satisfy (most of which refer to photospheric conditions), as can be found in recent studies \citep[][]{2022_DalTio,2023_Massey,2019_yang}.
The development of asteroseismology techniques, reviewed by \cite{Aerts_Mathis_2019}, can provide an insight into the internal characteristics of stars that also helps constrain stellar modelling \citep[e.g.][]{2023_burssens,2023_bowman,2019_eggenberger_a,2019_eggenberger_b}. 
Furthermore, the observations at ever larger redshifts provided by the JWST give hints about the early stellar generations in the universe that may have produced, for example, the seeds of supermassive black holes or strongly nitrogen enriched regions \citep[see for instance][]{2024_Ji,2024_Jeon, Nandal2024, Nandal2024b}.
Among the most prominent tasks for stellar models is to verify the constraints provided from all of these observations.

One major difficulty for massive stars above 8 $M_{\odot}$ is that the stellar core in the final phases of evolution, (i.e. from the beginning of carbon fusion in the core), is obscured from view deep inside the large stellar envelopes. The impact of mixing and nuclear burning in the stellar interior therefore cannot be well constrained by observations. Understanding the evolution of the stellar core in this regime heavily relies on numerical models, whose validation chiefly depends on observations of the subsequent evolution. Specifically, pre-supernova models are further employed as the starting points to predict the fate after core bounce and through to the supernova explosions \citep[e.g.][]{Ugliano_2012,Sukhbold_2014ApJ...783...10, Sukhbold_2016ApJ...821...38, Sukhbold_2018ApJ...860...93} as well as the multi-messenger signal of the post-explosion evolution \citep[e.g][]{Murphy:2009dx,Yakunin:2010fn,Scheidegger:2010en,Mueller:2012sv,Lund:2012vm,Cerda-Duran:2013swa,Tamborra:2013laa,Tamborra:2014hga,Tamborra:2014aua,Kuroda:2015bta,Mirizzi:2015eza,2016_kashiyama,Janka:2016fox,Andresen:2016pdt,Kuroda:2017trn,Janka:2017vlw,Kuroda:2017trn,Pan:2017tpk,OConnor:2018tuw,Powell:2018isq,Radice:2018usf,Morozova:2018glm,Walk:2018gaw,Andresen:2018aom,Lin:2019wwm,Walk:2019ier,Muller:2019upo,Andersen:2021vzo,Powell:2020cpg,Pan:2020idl,Mezzacappa:2020lsn,2021_schneider,Andersen:2021vzo,Jardine:2021fsf,Mezzacappa:2022xmf,Bugli:2022mlq,2023_bugli,Vartanyan:2023sxm}. With these predictions, we may compare observations of the observed statistics of neutron stars and black holes, including constraints on their spins, supernova explosion properties, predicted gravitational waveforms, neutrinos or nucleosynthetic yields with the obtained results, and hence constrain the pre-supernova models \citep[e.g.][]{2009_smartt,2015_smartt,2019_radice,2023_fonseca,2024_KamLAND,Reichert_2021MNRAS.501.5733,Reichert_2023MNRAS.518.1557,Reichert_2024MNRAS.529.3197,2023_Sieverding}.

Late-phase stellar evolution models have been studied by numerous authors using a variety of different stellar evolution codes. With  GENEC, \cite{Hirschi_2004} and \cite{Ekstrom_Georgy_Eggenberger_Meynet_Mowlavi_Wyttenbach_Granada_Decressin_Hirschi_Frischknecht_et_al._2012}, pushed models beyond silicon-core burning. For the KEPLER stellar evolution code studies such as \cite{Heger_2001PhRvL..86.1678},\cite{Sukhbold_2014ApJ...783...10}, and \cite{Sukhbold_2018ApJ...860...93} have produced a large number of pre-supernova models. The development of MESA in \cite{Paxton_Bildsten_Dotter_Herwig_Lesaffre_Timmes_2011,Paxton_Cantiello_Arras_Bildsten_Brown_Dotter_Mankovich_Montgomery_Stello_Timmes_et_al._2013,Paxton_Marchant_Schwab_Bauer_Bildsten_Cantiello_Dessart_Farmer_Hu_Langer_et_al._2015,Paxton_2018,Paxton_Smolec_Schwab_Gautschy_Bildsten_Cantiello_Dotter_Farmer_Goldberg_Jermyn_et_al._2019} has also produced grids of pre-supernova models.  All these studies have provided initial models to probe the subsequent dynamical evolution of collapsing massive stars. 

The properties of progenitor stars at the moment of collapse are crucial in order to determine the outcome of the post-collapse evolution of massive stars. Among these properties we mention, the density, the rotational and magnetic profiles, the mass of the iron core and its radius \citep[ordinarily quantified in terms of its compactness parameter][]{OConnor_Ott_2011ApJ...730...70}, or its electron fraction profile \citep[e.g.][among others]{Baron_Cooperstein_1990ApJ...353..597}. Supernova explosions are very sensitive to the progenitor structure; the entropy, electron fraction, equation of state, and neutrino physics all have a strong impact \citep{1984_Cooperstein,1985_baron,1990_Bethe}. The stellar profiles of these quantities result from a complex balance of nuclear burning stages and the creation, spatial dimension, and temporal sequence of different convective layers (some of which may mutually interact). Convection in one-dimensional stellar evolution codes is implemented assembling together the Schwarzchild-Ledoux (local) instability criterion (either the Schwarzshild or the Ledoux criterion is used, but not together), and basic prescriptions for diffusion of chemical species and angular momentum (in addition to parametric models for semi-convection, overshooting). Since the incorporation of these elements varies from one stellar evolution code to another, more often than desirable, the pre-supernova state for seemingly similar zero age main sequence (ZAMS) conditions differs among different groups \citep{Jones_Hirschi_Pignatari_Heger_Georgy_Nishimura_Fryer_Herwig_2015,Chieffi_Limongi_2020ApJ...890...43}. 

In this paper, we show the details of the implementation of an extended equation of state (EoS) in GENEC covering the whole density-temperature plane required to reach core collapse for massive stars. This EoS patches together two pre-existing EoSs, which include complementary physical aspects, along with some common elements (Sect.\,\ref{sec:neweos}). We show that while the local values of the EoS recovered thermodynamic variables do not show large discrepancies, their derivatives may have significant deviations, specifically derivatives impacting the criteria used to distinguish among zones that are unstable or stable to convection. Additionally, we extend the coverage of the radiative and conductive opacities to high-temperatures and densities to include the effects of degeneracy and high-energy physics, which play an important role in the final phases of stellar life (Sect.\,\ref{Sec:opacities}). With the aim of approximately covering the most advanced burning phases of the evolution of massive stars, we also extend the existing reduced nuclear reaction network to include electron capture on heavy nuclei (Sect.\,\ref{sec:Network}). Therefore, the nuclear network approximately accounts for the reduction of the electron fraction ($Y_e$) and consequently the electron pressure, which induces core collapse. To validate these improvements, using GENEC, we evolved several models of various massive stars until the brink of core collapse, which we compare to corresponding models evolved with MESA (Sect.\,\ref{sec:results}). We discuss our results and present our conclusions in Sect.\,\ref{sec:discussion}. Since the implementation details of the previous default EoS in GENEC are not easily accessible in electronic form, we highlight the most relevant features of that EoS in the Appendix~\ref{sec:Dichte_EOS}.

\section{A new EoS to cover the regime from partial ionisation to relativistic electron degeneration}
\label{sec:neweos}

Stellar matter is usually modelled as a mixture of ideal (baryonic) gases, radiation, and electrons/positrons. It encompasses regimes of partial to full ionisation. The current version of the GENEC stellar evolution code employs, by default, the Dichte\footnote{The name Dichte comes from the German word for density as the code was developed within the group of Kippenhahn and the EoS returns the density when one inputs pressure, temperature, and composition.} EoS (D-EoS, hereafter), which accounts for the partial ionisation regime of certain light elements, (hydrogen, helium, carbon, oxygen, nitrogen, and magnesium) in the stellar envelope (see Appendix~\ref{sec:Dichte_EOS} for details), as well as fully ionised medium of the stellar interior. We note that extending the treatment of partial ionisation for heavier elements would be welcome, as discussed in \cite{Jermyn_2021ApJ...913...72}, but this is a matter for future work. 
Two important regimes for massive star evolution are not covered by the D-EoS, namely the electron degeneracy in the high-density (i.e. relativistic) regime and the effects of electron-positron pair creation in the very high-temperature regime. To cover this region of the density-temperature parameter space, we have extended the existing GENEC EoS by matching it with the Timmes EoS (T-EoS, hereafter) \citep{Timmes_2000} in the deep stellar interiors. The latter EoS 
%, as well as including the ionic and radiative contributions,
computes the electron-positron contribution based upon an interpolation table of the Helmholtz free energy (ensuring thermodynamic consistency). 
The T-EOS also includes the Coulomb energy term, absent from the D-EoS, which can be relevant in high-density and low-temperature conditions, however, such conditions are generally not reached in the models presented in this work.
\footnote{The implementation of the T-EOS in GENEC, employs the subroutines read\_helm\_table, read\_helm\_iontable, and helmeos from the freely available version on the web at https://cococubed.com/codes/eos/helmholtz.tbz, version of 2022-01-08.} It should be noted that the Timmes EoS (T-EoS hereafter) assumes total ionisation of the plasma, which is not the case in the whole stellar interior. To allow for the EoS to deal with the partial ionisation in the stellar envelope, we employ a switch between the two EoSs. According to \cite{Jermyn_2021ApJ...913...72} the ionisation effects become relevant for temperatures $T<T^\text{ion}\approx 10^4Z_j^2\,$K and densities $\rho<\rho^\text{ion}\approx 3A_jZ_j^3\,\text{g\,cm}^{-3}$, where $Z_j$ and $A_j$ are the atomic number and atomic mass of the considered element. Using these criteria on the ionisation conditions as approximate guidelines, and trying to minimise the jump between the two different EoSs, if the following condition is verified, we use the D-EoS
\begin{equation}
\label{eqn:EOS_cond}
    \hspace{2.3cm} \rm log(\rho) \leq 2.8 \ \ \rm{or} \ \ \rm log(T) \leq 7.55,
\end{equation}
otherwise the T-EoS is called. This switch is similar to the matching point used in MESA between EoS tables and the same T-EoS \cite[see Fig.\,1 of][]{Paxton_Bildsten_Dotter_Herwig_Lesaffre_Timmes_2011}. The matching conditions also ensure that the species considered in the partial ionisation calculation are, at this point, fully ionised. For the most abundant species in the stellar envelope, hydrogen and helium, this approximately happens for densities and temperatures above 
\begin{align}
    (\rho_\text{ion}^{\rm H},T^\text{ion}_H) &\approx (3 \ \text{g\,cm}^{-3},10^4\ \text{K}), \\
    (\rho_\text{ion}^{\rm He},T^\text{ion}_{\rm He}) &\approx (10^2 \ \text{g\,cm}^{-3},4\times10^4 \ \text{K}).
\end{align}
Thus, noting that the switch to full ionisation is handled according to Eq.\,\eqref{eq:ionizationbypressure} in the D-EoS, the partial ionisation regime for hydrogen and helium is entirely addressed by this EoS (and not by the T-EoS). In the following, we collectively refer to the coupled EoS employed in GENEC as the DT-EoS.

\subsection{Electronic contribution to the pressure}
The T-EoS represents an improvement over the D-EoS when it comes to computing the electronic pressure (see Appendix~\ref{sec:Dichte_EOS}). While the D-EoS employs analytic estimations for the evaluation of the Fermi-Dirac integrals that are used to compute the electron contribution to the thermodynamic variables, the T-EoS uses a more accurate table for the leptonic contribution including electrons and positrons. 

\begin{figure}[ht]
    \centering
    \includegraphics[width=0.5\textwidth]{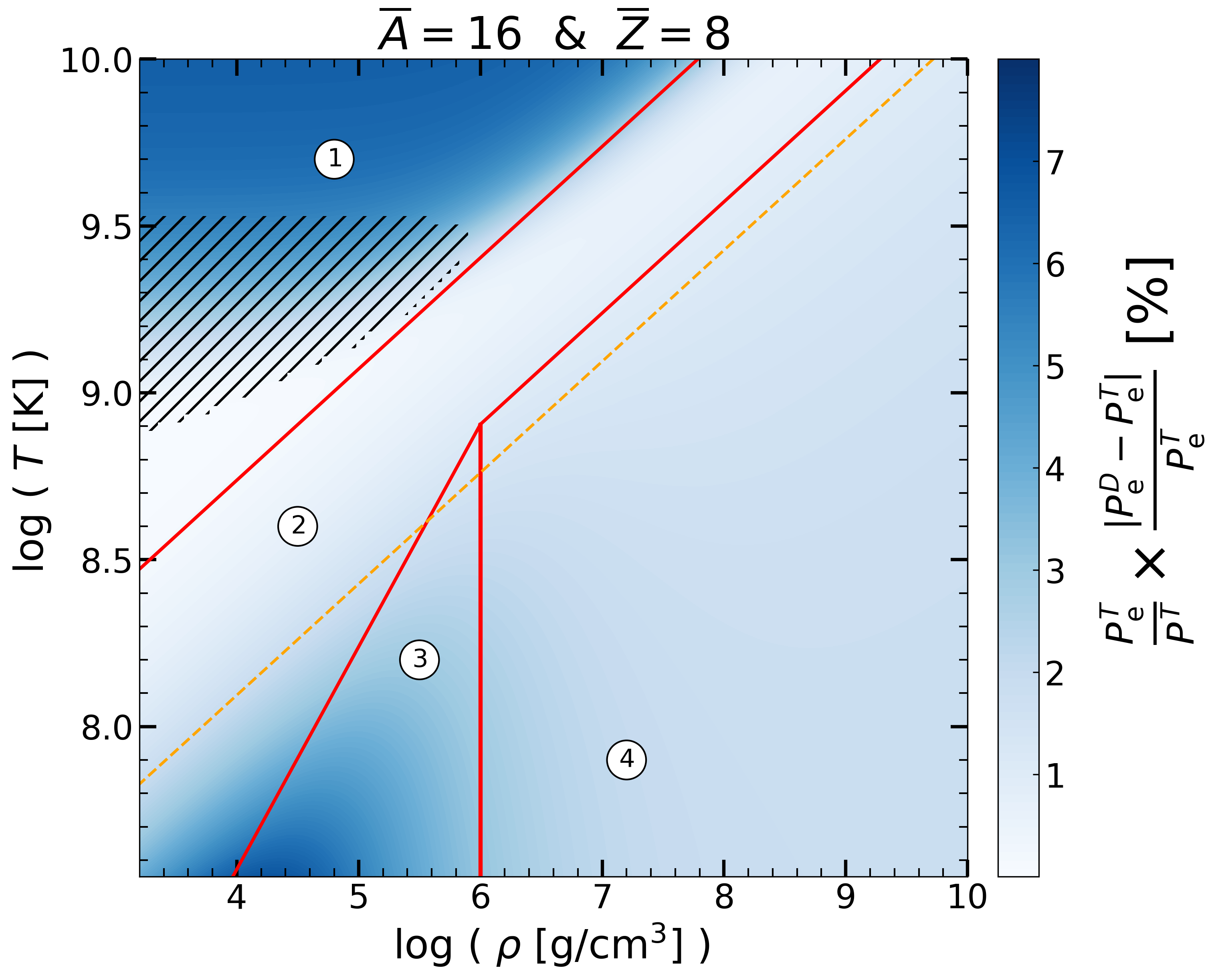}
    \caption{Variation between the lepton pressure computed using the D-EoS with that computed by the T-EoS. The EoSs are evaluated for a pure $\rm ^{16}O$ composition, $\overline{A}=16$ and $\overline{Z}=8$. We also highlight four different regions separated by the red lines. Zone \Circled{1} corresponds to the region where the total pressure is dominated by radiation pressure. In zone \Circled{2} perfect gas pressure is dominant. Zone \Circled{3} is dominated by non-relativistic degenerate electron pressure.  Zone \Circled{4} marks the limit where the electrons become relativistic. The region where $\Gamma =\left. \frac{\partial \ln P}{\partial \ln \rho}\right|_{\rm ad} < 4/3$ is black-hashed. 
    Finally the orange dashed line corresponds to the relation $T=2.27\times10^5\rho^{1/3}\overline{Z}^2/\overline{A}^{1/3} $, below which $\Gamma_{\rm C}\le 1$ (Eq.~\eqref{eq:Gamma_C}) and Coulomb corrections to the EoS are important.}
    \label{fig:P_diff}%
\end{figure}

In Fig.~\ref{fig:P_diff} we show the variations of the electron-positron pressure contribution as a function of  $\rho$, and  $T$,  between the two EoSs. The part of the $(\rho,T)$-plane represented in that figure falls fully inside the region where we employ the T-EoS. For the computation of the pressure we take the case of pure $\rm ^{16}O$, for which $\overline{A}=16$ and $\overline{Z}=8$, although changing the composition does not alter the conclusions made here. The quantity represented in Fig.~\ref{fig:P_diff} is
\begin{equation}
\label{eqn:P_difference}
    \chi = \frac{P_{\rm e}^{T}}{P^{T}} \times \frac{\left| P_{\rm e}^{D}-P_{\rm e}^{T}\right| }{P_{\rm e}^{T}},
\end{equation}
where $P_{\rm e}^{T}$, $P_{\rm e}^{D}$ are the electronic pressure contributions of each EoS, and $P^{T}$ is the total pressure of the T-EoS. We weight by the factor $P_{\rm e}^{T}/P^{T}$ to highlight regions where the electronic pressure is a relevant contribution to the total pressure. The main difference between each EoS is the presence of the electron-positron calculation in the T-EoS, along with the treatment of the Coulomb energy term (also related to the electronic pressure), thus the difference in pressure between the two is primarily relevant in the electronic pressure. The two EoSs concur for the electronic pressure in most of the $(\rho,T)$ plane. The pressure differences are only of a few percent, reaching 7\,\% at maximum. The largest differences are in 2 regions, the upper left corner of Fig.~\ref{fig:P_diff}, where the pair creation and the impact of the positrons are significant, and the transition between zone \Circled{2} to \Circled{3}, where the analytic approximations of the Fermi-Dirac integrals used by the D-EoS depart from the tabulated version of the T-EoS. Also, in this region, the Coulomb corrections (included in the T-EoS, but not in the D-EoS) contribute to the differences. To see more clearly where these corrections have an effect, we display in Fig.~\ref{fig:P_diff} (orange dashed line) the line corresponding to a ratio of Coulomb to thermal energy of the ions
\begin{align}
\label{eq:Gamma_C}
\Gamma_{\rm C}=2.27\times10^5\frac{\rho^{1/3}}{T} \frac{\overline{Z}^{\,2}}{\overline{A}^{\,1/3}}
\end{align}
equal to unity. Below this line Coulomb corrections contribute significantly to the EoS \cite[see, e.g.][Eq.~(7.120) for further details]{Maeder_2009}.

\subsection{ Impact on density }
\begin{figure}[ht]
    \centering
    \includegraphics[width=0.5\textwidth]{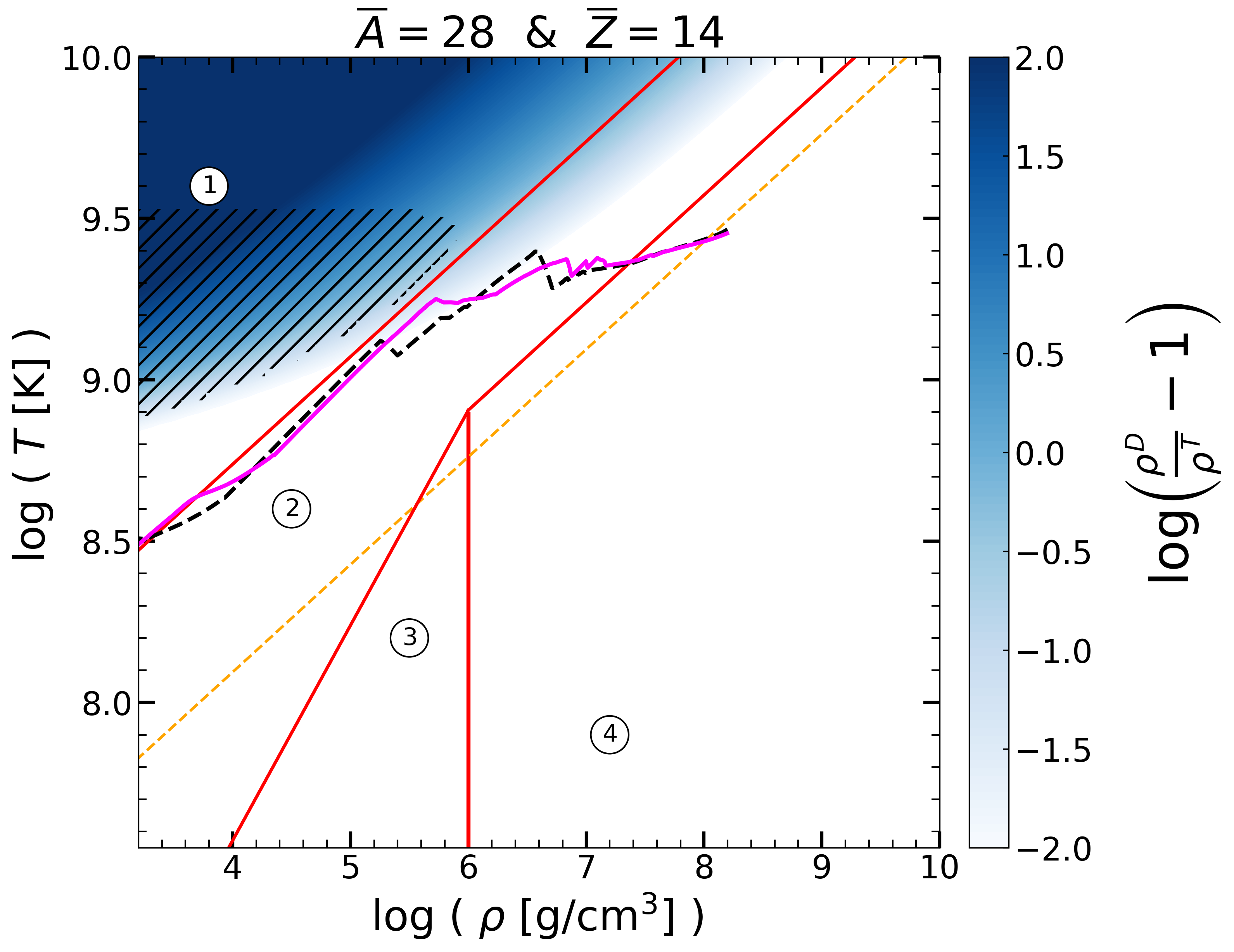}
    \caption{Comparison of the density from T-EoS, $\rho^{\rm T}$, and the output density computed with the D-EoS, $\rho^{\rm D}$. The composition used is that of pure silicon, $\overline{A}=28$ and $\overline{Z}=14$. White regions denote that $\rho^{\rm D}$ is very close to $\rho^{\rm T}$. Blue regions indicate that $\rho^{\rm D}\ \gg\ \rho^{\rm T}$. Shown are the same numbered zones as in Fig.~\ref{fig:P_diff} and the same hatched region. In  dashed black is the internal $(\rho(r),T(r))$ structure of a $15 M_{\odot}$ model at the onset of silicon core burning, computed using the DT-EoS and in magenta the same model using the D-EoS.} %
    \label{fig:rho_with_25M}%
\end{figure}

The change of EoS also impacts the density profile in the star, a crucial element for the determination of the stellar structure. The T-EoS takes as input a pair of values $(\rho,T)$ to construct the Helmholtz free energy and evaluate the pressure $P^{\rm helm}$ \citep{Timmes_Swesty_2000}. On the other hand, the D-EoS takes as input the pair of values $(P,T)$ and iterates over the degeneracy parameter, $\Psi$, or electron chemical potential,\footnote{We recall that the degeneracy parameter is the electron chemical potential in units of $k_\mathrm{B}T$, where $k_\mathrm{B}$ is the Boltzmann constant.} to find the density ($\rho^{\rm D}$) that satisfies the equations shown in the Appendix~\ref{sec:Dichte_EOS}.  In Fig.~\ref{fig:rho_with_25M} we show the logarithm of the difference $ \rho^{\rm D}/\rho^{\rm T} - 1 $. Where $\rho^{\rm T}$ is the input density for the T-EoS and $\rho^{\rm D}$ the output density of the D-EoS. $\rho^{\rm D}$ is obtained by using as input value $P^{\rm T}$ (i.e.  the pressure computed using the T-EoS for the thermodynamic conditions $(\rho^{\rm T},T)$). We note that the temperature, $T$, has been fixed in all the process of comparison.

We see in Fig.~\ref{fig:rho_with_25M} that in a large part of the $(\rho,T)$ plane $ \log ( \rho^{\rm D}/\rho^{\rm T} - 1 ) < -2$. It is only in the upper left region, where the pressure is dominated mainly by the radiation contribution, that the D-EoS begins to diverge strongly from the thermodynamic consistent T-EoS. In this region, the contributions made by the positrons to the pressure can be large. In general, the presence of positrons in the EoS will affect the electron chemical potential and thus the degeneracy parameter. The lack of positrons in the D-EoS, and the analytic expressions used to evaluate the Fermi-Dirac integrals (Appendix \ref{sec:Dichte_EOS}), leads to large differences in the estimation of the degeneracy parameter between the two EoSs. For example, at the point $(\rho,T) = (10^5\ \rm g\,cm^{-3},10^{9.5} \ \rm K)$, for a pure silicon composition, the T-EoS returns $\psi^T=-1.8$ whereas the D-EoS returns $\psi^D=-1.2$. This degeneracy parameter with the D-EoS gives a density of roughly $6\times 10^5\, \rm g\,cm^{-3}$ a considerable change to the Timmes input of $1\times 10^5\, \rm g\,cm^{-3}.$ Thus using the Timmes EoS (or more sophisticated alternatives; see e.g. \citealt{Jermyn_2021ApJ...913...72}) in this region is mandatory to obtain a correct evaluation of the density and, thereby, the correct stellar structure.

Most of the stellar interior of our models are not affected by this region of strong differences in the density. However, in Fig.~\ref{fig:rho_with_25M} we super-impose the $(\rho,T)$ internal structure of a $15M_{\odot}$ model computed with the D-EoS in magenta and a model using the DT-EoS. The main shift between these models comes from the location of the interfaces, represented by sharp variations in the $(\rho,T)$ curve. The size of jump at these interfaces in density and temperature is also different between the two models using different EoSs. The properties at shell interfaces are as we see in Sect.~\ref{sec:discussion} crucial for understanding supernova explosions of these progenitors.

\subsection{Impact on thermodynamic derivatives}
\begin{figure}[htb!]
    \centering
    \includegraphics[width=0.5\textwidth]{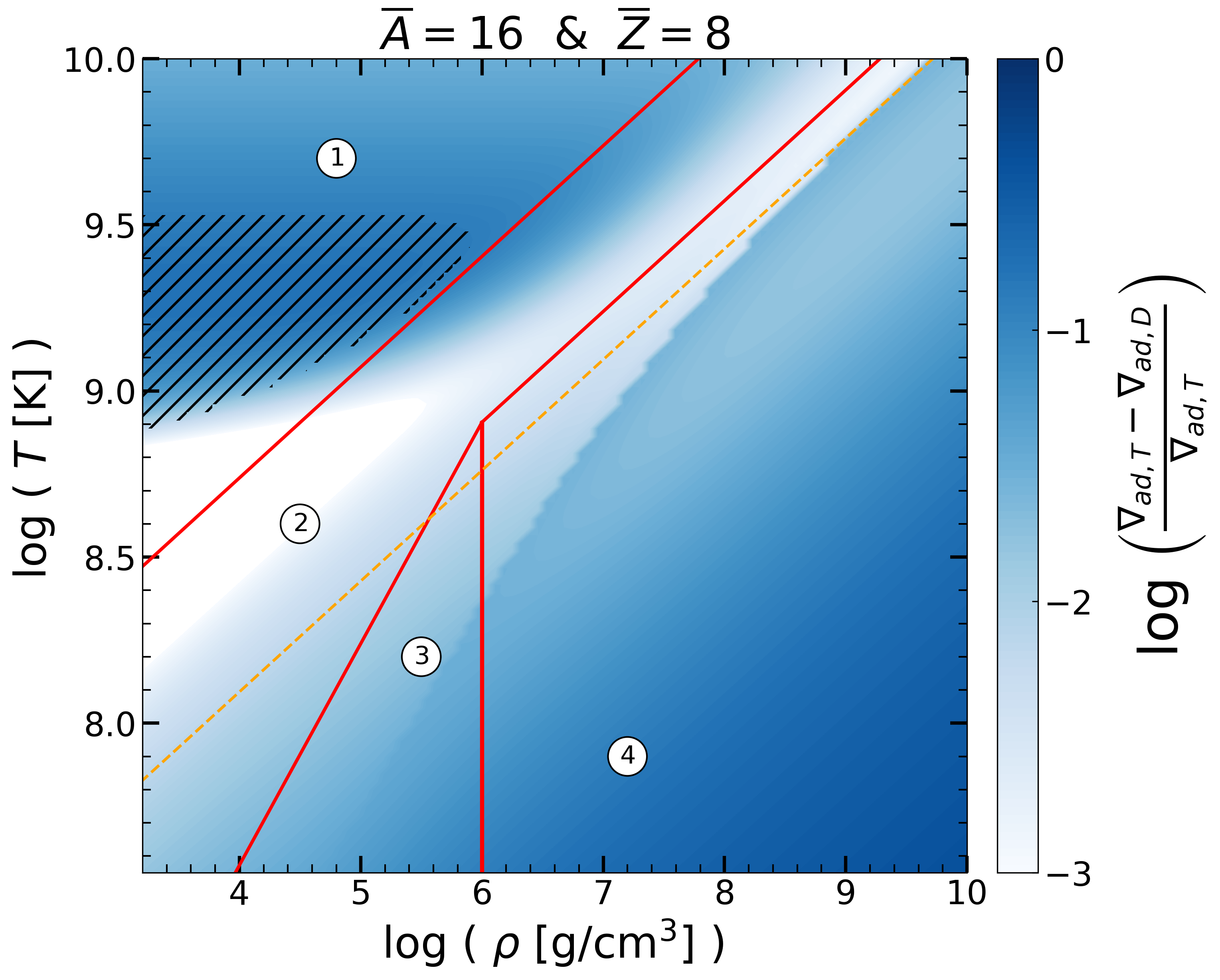}
    \caption{ Relative difference of the thermodynamic derivative $\nabla_{\rm ad}$ between the two EoSs, D-EoS and T-EOS, in log scale. The composition used is that of pure $\rm ^{16}O$, $\overline{A}=16$ and $\overline{Z}=8$. The numbered zones represented are the same as in Figs.~\ref{fig:P_diff} and ~\ref{fig:rho_with_25M}.}%
    \label{fig:Nabla_ad}.%
\end{figure}
 Between different EoSs the thermodynamic derivatives can change significantly. Specifically, we focus on the derivative needed to compute the adiabatic gradient:
\begin{equation}
\label{eqn:nablad_def}
    \nabla_{\rm ad} \equiv \left( \frac{\rm d \ln T }{\rm d \ln P} \right)_{\rm ad}.
\end{equation} 
This must be obtained with sufficient accuracy, since $\nabla_{\rm ad}$ is a fundamental variable to determine whether a mass shell is unstable to convection. 

In Fig.~\ref{fig:Nabla_ad} we show the relative difference between the adiabatic gradients computed with the two EoSs. Most massive star models will have most of their mass under the conditions of zone \Circled{2} of the $(\rho,T)$ plane, where the differences are minimal. However, we see that in the upper region of zone \Circled{2}, close to the border with zone \Circled{1}, the difference begins to increase. These conditions are reached by the inner regions of massive stars and, thus, the practical consequence of the difference in the computed values of $\nabla_{\rm ad}$ will be the modification of the size of the convective core or of convective shells surrounding it. The sensitivity of stellar modelling to convective core size and problems of overshooting is well known, accurate estimations of convective instability is essential in stellar modelling.

\section{Extension of the opacity coverage}
\label{Sec:opacities}
The current implementation of opacities for the advanced phases in GENEC does not cover the full $(\rho,T)$ plane needed to evolve massive star models until  core collapse. We shall use analytic extensions to estimate the opacities in these high-energy and relativistic regimes. 

When computing the total flux of energy, the radiative and conductive flux sum up to give a total energy flux. This is given by,
\begin{equation}
    F_{\rm tot} = F_{\rm rad}+F_{\rm cond} = (C_{\rm rad}+C_{\rm cond})\frac{dT}{dr},
\end{equation}
where $C_{i} = 4acT^3/(3\kappa_{i}\rho)$. As the fluxes are proportional to the inverse of the respective opacities,  the total opacity is,
\begin{equation}
    \frac{1}{\kappa} = \frac{1}{\kappa_{\rm cond}} + \frac{1}{\kappa_{\rm rad}}.
\end{equation}
In the following we detail the extensions employed to compute both $\kappa_{\rm cond}$ and $\kappa_{\rm rad}$ in the high-density and high-temperature regime, which is reached as stellar models evolve to core-collapse.

\subsection{Radiative opacity}

For most of the $(\rho,T)$ plane the radiative opacities are evaluated from tabulated values \citep[see][for further details]{Ekstrom_Georgy_Eggenberger_Meynet_Mowlavi_Wyttenbach_Granada_Decressin_Hirschi_Frischknecht_et_al._2012}, which take into account many physical processes such as bound-bound transitions, bound-free transitions, free-free transitions and electron scattering to name just a few. The coverage of these tables generally only extend to temperatures of $10^9$\,K. At higher temperatures, the opacity is dominated by electron scattering, but the Klein-Nishina corrections to Compton scattering lower it \citep{Weaver_Zimmerman_Woosley_1978}, and at even higher temperatures, the effects of electron-positron pairs must be included in the opacity \citep{Woosley_Heger_Weaver_2002}. In the high-density regime, electron conduction is an important contribution to the opacity \citep[see][]{Itoh_1985ApJ...294...17}, and also electron degeneracy tends to reduce the opacity due to occupation of the electron phase space. To extend the computation of the radiative opacity to these regions, we apply the work of \cite{Poutanen_2017} to estimate the Rosseland mean opacity for electron scattering in the high-temperature regime. This opacity is expressed as
\begin{equation}
\label{eqn:kappa_es}
    \kappa_{\rm es} = \frac{\sigma_T N_e}{\rho} \frac{1}{\Lambda},
\end{equation}
where the parameter $\Lambda(T,\psi)$ is obtained from a fit in temperature and degeneracy parameter. The fit employed is,
\begin{equation}
\label{eqn:Lambda}
    \Lambda(T,\psi) = f_1(\psi) \left[1 + (T/T_{\rm br})^\alpha\right],
\end{equation}
where,
\begin{align}
\centering
    T_{\rm br} &= T_0 f_2(\psi), \\
    \alpha &= \alpha_0 f_3(\psi), \\
    f_i(\psi) &= 1 +c_{i1}\xi+c_{i2}\xi^2, \hspace{0.5cm} i=1,2,3 \\
    \xi &= \exp(c_{01}\psi +c_{02}\psi^2).
\end{align}
The fitting parameters chosen are for the range $2-300\,$keV and listed in Table.~\ref{tab:coeffs}.

\begin{table}
\centering
\caption{Coefficients used to compute the expression of $\Lambda(T,\Psi)$, Eq.~\ref{eqn:Lambda}. Taken from \cite{Poutanen_2017} for the range 2-300\,keV.}
    \begin{tabular}{||c c||}
    \hline
    Coefficient &  2-300 keV \\[0.5ex]
    \hline
    \hline
         $T_0$ & 43.3 \\
         $\alpha_0$ & 0.885 \\ 
         $c_{01}$ & 0.682 \\ 
         $c_{02}$ & -0.0454 \\ 
         $c_{11}$ & 0.24 \\ 
         $c_{12}$ & 0.0043 \\ 
         $c_{21}$ & 0.050 \\ 
         $c_{22}$ & -0.0067 \\ 
         $c_{31}$ & -0.037 \\ 
         $c_{32}$ & 0.0031 \\[0.5ex] 
    \hline
    %\hline
    \end{tabular}

    \label{tab:coeffs}
\end{table}

\subsection{Conductive opacity}

The conductive opacities due to electrons are computed using the formulation of \cite{Iben_1975}, given in the Appendix of that paper. These were already employed in the previous version of GENEC \citep{Ekstrom_Georgy_Eggenberger_Meynet_Mowlavi_Wyttenbach_Granada_Decressin_Hirschi_Frischknecht_et_al._2012};  however, here  the expressions have been modified to apply also in the final phases of stellar evolution. For this reason, we detail next the procedure used to compute $\kappa_{\rm cond}$. There are two main cases, the first being when the electrons are non-relativistic, roughly for $\rho < 10^6\,\text{g\,cm}^{-3}$. In that case, a fit is developed based on the results of \cite{Hubbard_Lampe_1969}, where they compute the thermal electron conductivity for mixtures of pure hydrogen, helium, and carbon. Thus, for densities smaller than $10^6\,\text{g\,cm}^{-3}$ the conductive opacity is written as
\begin{equation}
    \kappa_{\rm cond} = (X\theta_X + Y \theta_Y + Z_C \theta_C ) / ( T_6 f )
\end{equation}
where X and Y are the mass fractions of hydrogen and helium, respectively, and
\begin{equation}
\label{eqn:Z_c}
    Z_C = \frac{1}{3} \sum_i Z_i^2 X_i / A_i
\end{equation}
for all species beyond helium, ($T_6 = T / 10^6$).\footnote{In the previous versions of GENEC, only a selection of all species above helium was included in $Z_C$. In particular, most of the species with $A\ge 56$, were not included. In the upgraded version, all species consistent with the nuclear reaction network used are for the computation of $Z_C$. The same comment applies to the quantities $Z_\alpha$ and $Z_\beta$ needed for the evaluation of $\kappa_{\rm cond}$ in Eq.~\eqref{eq:kappacond}.}\label{foot:kappac} Here $\theta_X$, $\theta_Y$, $\theta_C$, and $f$ are fit parameters, dependent on thermodynamic quantities, as detailed in  \cite{Iben_1975} to match the tabulated results of \cite{Hubbard_Lampe_1969}. We note that the factor 1/3 in Eq.~\ref{eqn:Z_c} is there to normalise the weight of $X_{\rm ^{12}C}$ to 1 ($6^2 / 12 = 3$) because the computation for $\kappa_{\rm cond}$ in \cite{Hubbard_Lampe_1969} is only done for pure Carbon. 

In the case of relativistic degenerate electrons, \cite{Iben_1975} proposes a fit for the results of \cite{Canuto_1970}, which extend up to densities of $\rho \approx 10^{12}\,\text{g\,cm}^{-3}$. The approximation is
\begin{equation}
\label{eq:kappacond}
    \kappa_{\rm cond} = 6.35 \times 10^{-8} \times \frac{T_6^2 Z_{\beta}}{\epsilon_F^{2.5}G} \times \frac{1}{(1+\epsilon_F)} \ \mathrm{cm^2\,g}^{-1},
\end{equation}
where
\begin{align*}
    \epsilon_F &= \left[1 + \left(\frac{\rho_6}{\mu_e}\right)^{2/3}\right]^{1/2} -1, \\
    Z_{\beta} &= \sum_i Z_i^2 X_i / A_i,\\
    \log G &=  [( 0.873 - 0.298 \log Z_{\alpha} ) ,\\ 
    &+ (0.333 - 0.168 \log Z_{\beta})M ] \times \left[1-(1+\gamma)^{-0.85}\right], \\
    Z_{\alpha} &= \sum_i Z_i^2 X_i / A_i^{1/3}, \\
    \gamma &= 22.76\left(\rho_6^{1/3}/T_6\right)Z_{\alpha} \\
    M &= \text{min}(1,0.5 +\log\epsilon_F).
\end{align*}
Here $\mu_e$ is the electron chemical potential and we use the convention that $\rho_6=\rho[\text{g\,cm}^{-3}]/(10^6\,\text{g\,cm}^{-3})$ and $T_6=T[\text{K}]/(10^6\,\text{K})$ are dimensionless quantities. These fits provide good coverage for large parts of the $(\rho,T)$ plane for generic compositions, and are generally accurate to the order of a few percent. In the latest version of the MESA code, the conductive opacities are computed from the tables of \cite{Cassisi_Potekhin_Pietrinferni_Catelan_Salaris_2007} which have been extended to reach, $10^{10}\, \rm K$ in temperature, and $10^{11.5} \, \text{g\,cm}^{-3}$ in density \citep[see][]{Paxton_Cantiello_Arras_Bildsten_Brown_Dotter_Mankovich_Montgomery_Stello_Timmes_et_al._2013}. 
For our purposes, the fits provided by \cite{Iben_1975} estimate to sufficient accuracy the influence of electron conduction in the total opacity.

\subsection{Impact of updated opacity}

\begin{figure}[ht]
    \centering
    \includegraphics[width=0.5\textwidth]{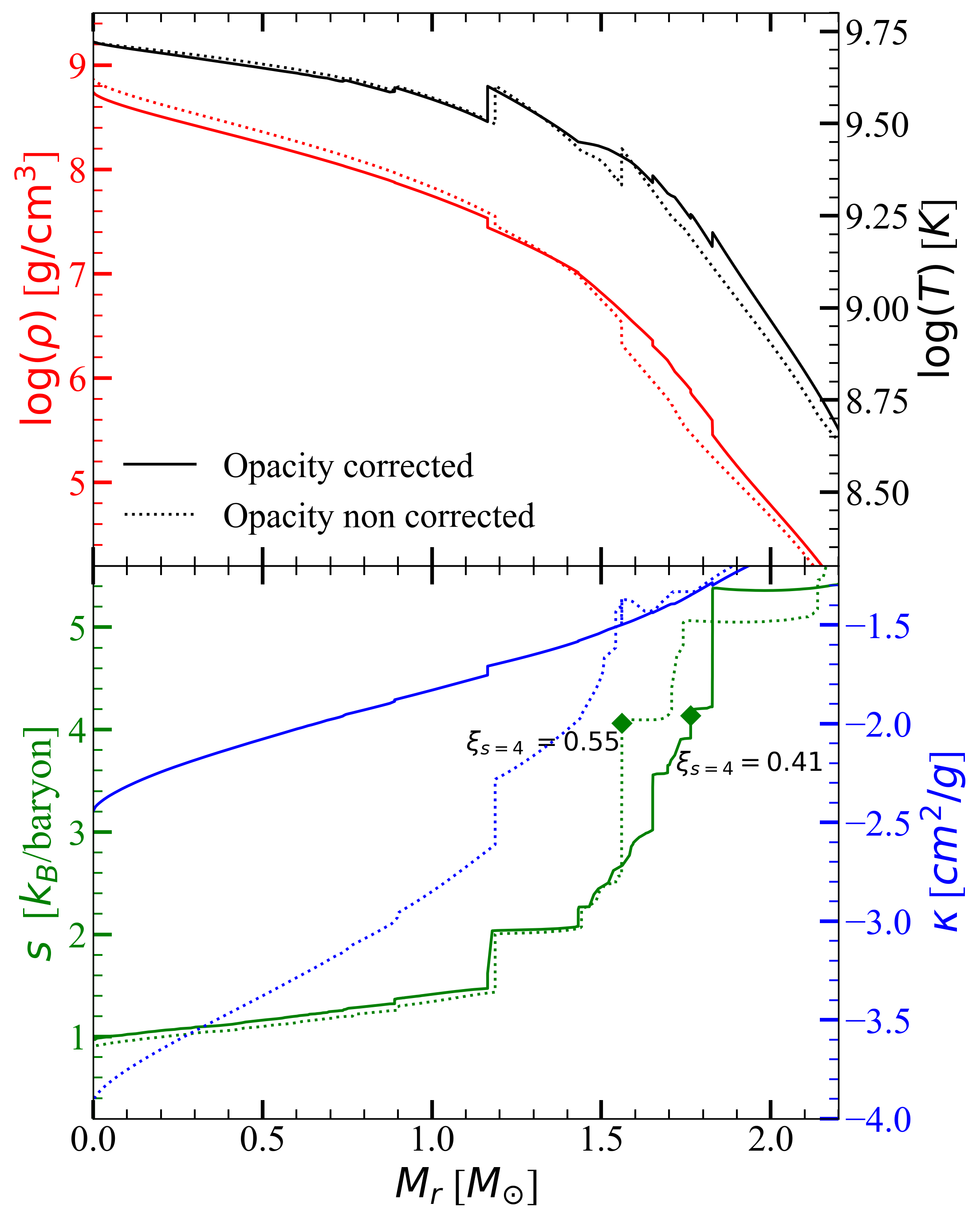}
    \caption{Thermodynamic variables plotted for $15M_{\odot}$, with and without the opacity corrections. The model is plotted during silicon shell burning some minutes before collapse. The solid line shows the model using the opacity upgrades for the radiative and conductive opacity as detailed in Sect.~\ref{Sec:opacities}. The dotted line shows the model run with the opacity previously implemented in GENEC. We display the compactness at an entropy per baryon $s=4$, computed using the definition in Eq.~\ref{eqn:compactness}.}%
    \label{fig:Opacity_compar}%
\end{figure}

To demonstrate the impact of the improved opacity calculation on the evolution of massive stars, we have computed a test model of initial mass $15M_{\odot}$ and solar metallicity, which is exactly the same as the model we present later in Sect.~\ref{sec:results}, except that the improved opacities of the previous sections have not been used. The new opacity routine only impacts the computation in the very final stages of evolution, where the density and temperature are high enough for the corrections to make any effect. In Fig.~\ref{fig:Opacity_compar}, we show the structure of the inner core for both of these models post-silicon core burning, and thus close to collapse.

The correction to the opacity is quite consequential for the final structure of the model, specifically, in the innermost regions of the star. That can be seen in Fig.~\ref{fig:Opacity_compar}, which displays the structure of the inner 2 solar masses for the $15M_{\odot}$ computed using GENEC with (solid lines) and without (dashed lines) the opacity corrections. The opacity changes by 2 to 3 orders of magnitude in the inner core. The most important contribution in this change is due to the inclusion of all species above helium (and not only a selection of them; see footnote~\ref{foot:kappac}) in the conductive opacity correction. Without the correction the GENEC model has a lower opacity and this leads to a more efficient cooling of the core. Therefore to rise the temperature to the same value the core must contract more. We see that the central density, shown in Fig.~\ref{fig:Opacity_compar} is slightly higher for the non corrected model for the same central temperature. As the core has contracted further the entropy jump represented at $s=4$ is at a smaller Lagrangian mass coordinate. This leads to a more compact core as parametrised by the compactness, $\xi_{s=4}$\footnote{See Sect.~\ref{sec:results} for more discussion on this parameter.} Eq.~\ref{eqn:compactness}. While the original opacity yields $\xi_{s=4}=0.55$, one obtains $\xi_{s=4}=0.41$ for the model with corrected opacities. The applied corrections are thus crucial to determine the most accurate progenitor structure at collapse.

\section{Implementation of an effective network}

\label{sec:Network}

To reach the brink of core collapse, massive stars progressively neutronise their cores, reducing the electron fraction, $Y_e$, and with it the supporting degenerate electron pressure. 
Any reaction network accounting for the most advanced burning phases, must include (at least, approximately) the weak reactions that reduce the core electron fraction (mostly, electron capture onto elements of the iron group). In the previous work by \cite{Hirschi_2004}, GENEC models were computed post-silicon core-burning and even post-silicon shell burning, thus bringing them very close to collapse models. However, the network employed did not account for any decrease in $Y_e$. To extend the use of GENEC models as supernova progenitors, we have implemented electron capture in a reduced network to model the $Y_e$ decrease seen in stellar cores prior to collapse. Electron capture is especially important from  silicon core burning to collapse  \citep{Woosley_Heger_Weaver_2002}. 
%(Fig. 13 and Fig. 20 of that paper). 
Ideally, one would include a large set of reactions and isotopes to resolve with precision the $Y_e$ evolution. For instance, \cite{Rauscher_2002} have employed a network of up to $2\,200$ isotopes, \cite{Farmer_2016} using MESA computed the pre-supernova structure of  models with various nuclear reaction networks of different sizes and \cite{2018_limongi_chieffi} used a network of roughly 300 isotopes.

\begin{figure}[ht]
    \centering
    \includegraphics[width=0.5\textwidth]{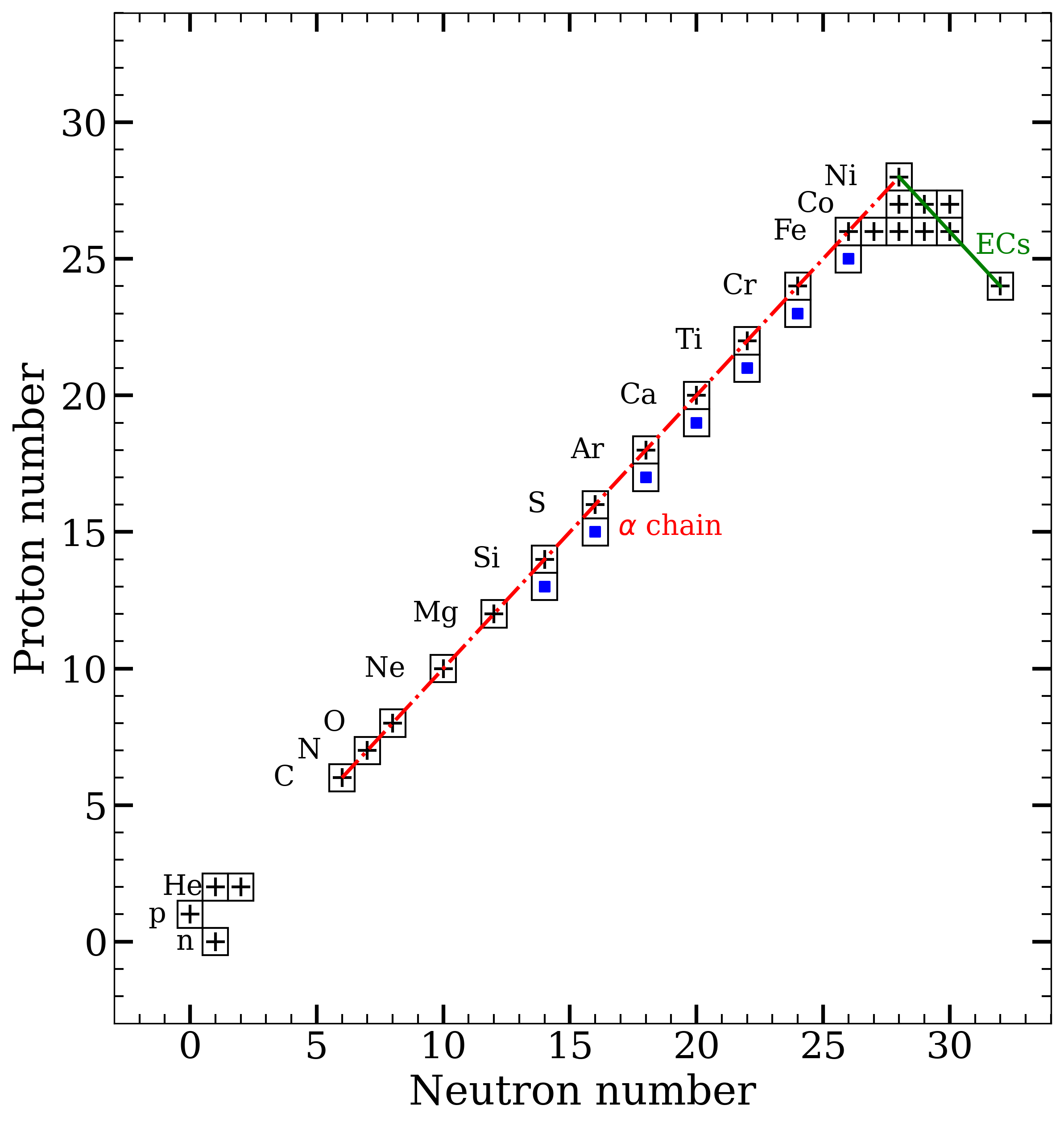}
    \caption{Network used in all GENEC models {from core C-burning onwards}. Squares filled with a plus sign denote isotopes explicitly evolved in the nuclear reaction network. Squares filled with a blue dot are isotopes implicitly included in the $\alpha$-chain. By this we mean that along the $\alpha$-chain are included not only the ($\alpha$, $\gamma$) reaction but also the ($\alpha$, $p$)($p$, $\gamma$) reactions without explicitly evolving the intermediate isotope abundance. With such a network, we get a good estimation of the energy generation rate \citep[see][for details]{Timmes_2000}. Additionally, compared to the previous GENEC network, we have the approximate electron capture from $\rm ^{56}Ni$ to $\rm ^{56}Cr$ to mimic pre-supernova $Y_e$ evolution. This network is very similar to that of the 22-isotope network of \cite{Farmer_2016}; we note that in Fig.~1 of that paper only 21 isotopes are flagged in the 22-isotope network, but $\rm ^{56}Co$ is added to the network making 22 in total. Compared with that network, we track explicitly extra iron group isotopes, namely $\rm ^{57}Co$ along with $\rm ^{53}Fe$ and $\rm ^{55}Fe$, and thus in total we track 25 isotopes explicitly.}%
    \label{fig:GeVal25}%
\end{figure}

In this work, we use a simple extension of the pre-existing $\alpha$-chain backbone network of GENEC. We approximate the effects of weak reactions that one would get in a larger network by taking one simple electron capture chain from $\rm ^{56}Ni$ to $\rm ^{56}Cr$. In this sense we create a similar network to the 21 species, detailed for example in \cite{Farmer_2016}, commonly refereed to in MESA literature \citep{Paxton_Bildsten_Dotter_Herwig_Lesaffre_Timmes_2011,Paxton_Cantiello_Arras_Bildsten_Brown_Dotter_Mankovich_Montgomery_Stello_Timmes_et_al._2013,Paxton_Marchant_Schwab_Bauer_Bildsten_Cantiello_Dessart_Farmer_Hu_Langer_et_al._2015,Paxton_2018,Paxton_Smolec_Schwab_Gautschy_Bildsten_Cantiello_Dotter_Farmer_Goldberg_Jermyn_et_al._2019} as approx21\_plus\_co56. The exact network we use contains 25 species\footnote{In the main sequence and helium core burning a larger network is used to capture carbon nitrogen oxygen, CNO, cycles correctly (see \cite{Ekstrom_Georgy_Eggenberger_Meynet_Mowlavi_Wyttenbach_Granada_Decressin_Hirschi_Frischknecht_et_al._2012} for details). From carbon burning onwards, we use the 25-species network described here.} and is presented in Fig.~\ref{fig:GeVal25}. Hereafter, we refer to this reduced network as GeValNet25. Compared with the approx21\_plus\_co56 network, we additionally include explicitly the iron group elements $\rm ^{53}Fe$, $\rm ^{55}Fe$, $\rm ^{55}Co$, and $\rm ^{57}Co$ to model the core neutronisation within the framework of GENEC.  

During $\rm ^{28}Si$ core burning, the composition of the core changes from elements grouped around silicon to isotopes grouped around nickel. Once the core is made up of elements in the iron group, further changes will happen along the rudimentary electron capture chain implemented here for species with $A=56$,

\begin{equation*}
    \hspace{1.5cm} \rm ^{56}Ni \ \xrightarrow\rm \ ^{56}Co \ \xrightarrow\rm \ ^{56}Fe \  \xrightarrow{(2)} \ \rm ^{56}Cr.
\end{equation*}
It should be noted that the last electron capture in this chain contains a double step, first from $\rm ^{56}Fe$ to $\rm ^{56}Mn$, and then to $\rm ^{56}Cr$. This electron chain mimics the effect of a much larger network containing many electron reactions \citep[cf. ][]{Aufderheide_1994ApJS...91..389,Langanke_Martinez-Pinedo_2000}. The rates used here were calibrated in a way that we reach similar core electron fractions as previous studies \citep{Weaver_Zimmerman_Woosley_1978, Woosley_Heger_Weaver_2002,Heger_Woosley_2010,Chieffi_Limongi_Straniero_1998,Farmer_2016,Paxton_2018}. It should be noted that direct beta reactions are not explicitly included in our calculations (differently from the inverse beta reactions, i.e. electron capture). Hence, there is no true balance between weak processes, implying that once the chain of electron captures begin, it unavoidably brings the core towards a composition dominated by the last isotope of the chain. In other words, should the core not collapse, in a finite time all the core would be composed of $\rm ^{56}Cr$. Thus, the electron capture rates have been suitably adapted to become effective weak rates that account for the state close to approximate beta equilibrium in the core.\footnote{Certainly, this is not a detailed beta equilibrium where every electron capture is balanced by the corresponding beta-decay, but rather, a global one, where the large values of the reaction rates allow that the integral of all beta decays approximately balances with the integral of all capture rates, facilitating  an effective equilibrium where $dY_e/dt$ is very small.}Admittedly, this approach introduces a simplification in modelling weak interaction physics, which facilitates the integration of these processes within the GENEC framework. Future updates could consider incorporating a selection of direct beta-decay reactions to approximate a global beta equilibrium. However, the sparse nature of the reaction network introduces uncertainties regarding the optimal sample of beta decays needed to accurately represent the underlying weak interaction processes. Energy changes resulting from effective electron capture are accounted for by applying the $Q$-value for these reactions, with an adjustment for the energy lost through neutrinos during weak interactions. The energy of the neutrinos is taken from the values found in \cite{Langanke_2001}. As pointed out in \cite{Heger_2001ApJ...560..307}, due to the quasi beta equilibrium in the core, the values of the beta decay rates are immaterial, as long as they facilitate the effective equilibrium (i.e. as long as they are sufficiently large). Even variations by factors 2 to 5 in the beta-decay rates do not significantly change the nuclear evolution. We note that the choice of using an electron capture chain along $A=56$ until $\rm ^{56}Cr$ is not by chance. In studies with larger networks and more species, the final $Y_e$ value commonly falls around $0.43-0.45$ \citep{Woosley_Heger_Weaver_2002}. If the simplified electron capture chain stopped at $\rm ^{56}_{26}Fe$, then the lowest $Y_e$ attainable would be $Y_{e}^{\rm min}= 26 / 56 \approx 0.46$ (when all the matter in a core mass shell would be converted to $\rm ^{56}_{26}Fe$), too high to match the results of larger and more complete networks. However, including $\rm ^{56}_{24}Cr$ in the chain allows one to evolve to $Y_{e}^{\rm min} = 24/56 \approx 0.42$. Thus, a core containing a mixture of $\rm ^{56}Fe$ and $\rm^{56}Cr$ can reach the electron fraction values generally found with sophisticated studies, even without explicitly accounting for the balancing effect of beta-decays. We cautiously point out that, with a reduced network, we may not provide a complete forecast of the nuclear composition at the brink of collapse.
We further note that, from silicon core burning onwards, the energy output in the core is not computed using each individual reaction but rather a quasi-statistical equilibrium (QSE) state between two elementary groups, around $^{28}\rm Si$ and $^{56}\rm Ni$. This is done by only taking into account the energy produced by reactions involving isotopes with $Z<14$, lower than $^{28}\mathrm{Si}$, and the $^{44}\rm Ti(\alpha,\gamma)^{48}\rm Cr$ reaction, which serves as the link between the 2 QSE groups of $^{28}\rm Si$ and $^{56}\rm Ni$ \citep[for further detail see][]{1998_Hix,Hirschi_2004}. This approximation is applied due to the lack of stability of the heyney solver during silicon core burning of the models when using the energy output of individual reactions. Coupling the entire chemical composition to the structure equations as done by MESA \citep{Paxton_Bildsten_Dotter_Herwig_Lesaffre_Timmes_2011, Paxton_Cantiello_Arras_Bildsten_Brown_Dotter_Mankovich_Montgomery_Stello_Timmes_et_al._2013, Paxton_2018, Paxton_Smolec_Schwab_Gautschy_Bildsten_Cantiello_Dotter_Farmer_Goldberg_Jermyn_et_al._2019} could likely create a more stable environment. Also, correctly including the feedback on the temperature within the nuclear network would help to evolve the energy generation coupled with the network. However, both of these points go beyond the scope of this work. The energy losses by the electron capture, carried away by neutrinos, are included explicitly on top of the QSE calculation, thus allowing for a consistent contraction of the core in the final phases, where electron capture is the dominant energy sink in the core for the stellar masses of interest. 
\begin{figure*}[ht]
    \centering
\includegraphics[width=\textwidth]{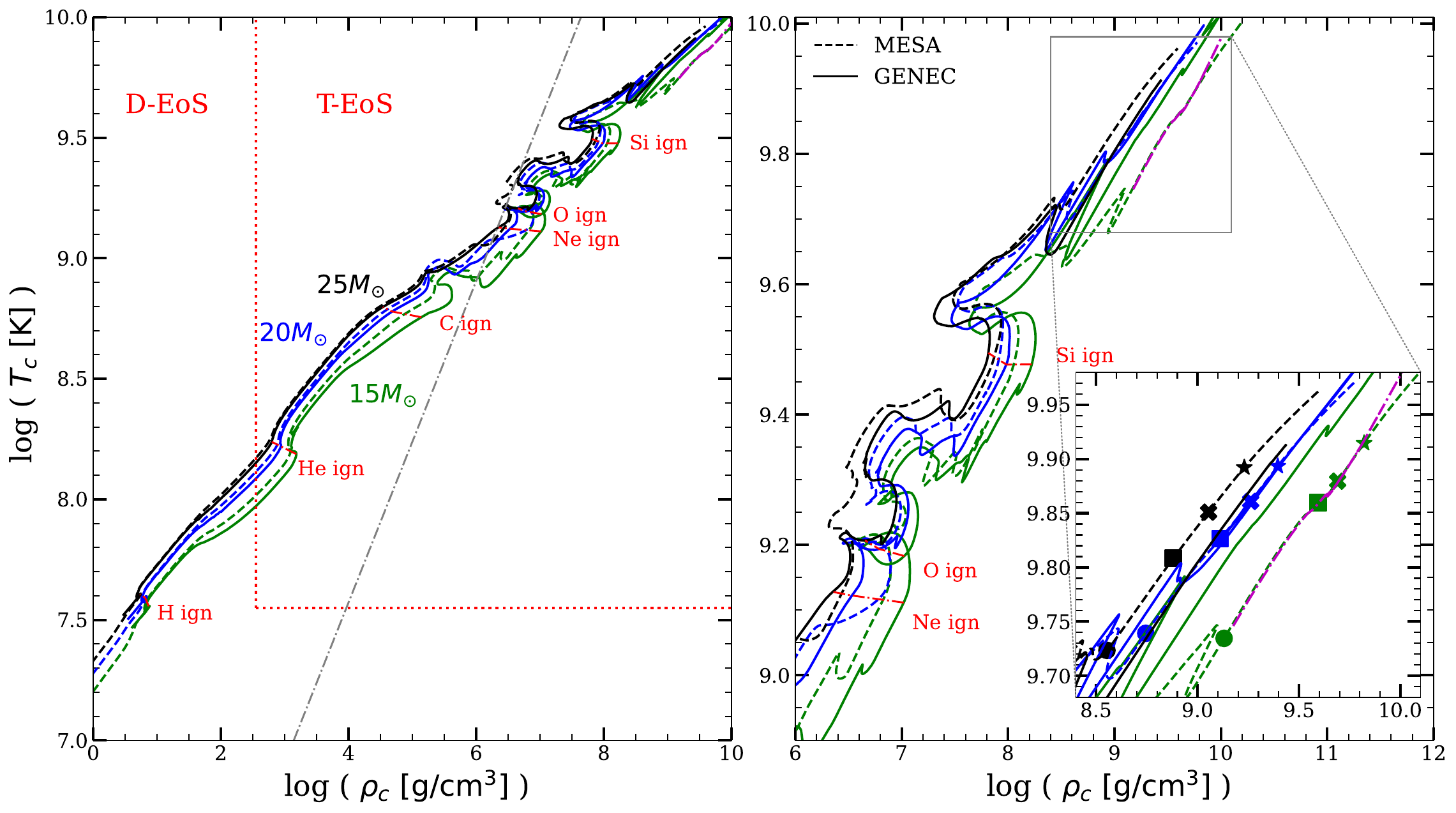}
   \caption{Evolution of central conditions for all models. Left: Evolution of the central temperature with respect to central density for all models. The GENEC models are shown as solid lined and the MESA models as  dashed. The $15M_{\odot}$ models are in green, $20M_{\odot}$ in blue, and $25M_{\odot}$ in black. The magenta dash-dotted line is a 15$M_{\odot}$ MESA model run without the acceleration term for comparison (only the very end of the evolution is shown because the prior evolution is identical to the other MESA model). The grey dash-dotted line shows the limit between non-degenerate and degenerate electron gas. The red dotted lines seperate the regime where the D-EoS is applied and when the T-EoS is used, as given by Eq.~\eqref{eqn:EOS_cond}. The beginning of each burning phase is indicated. Right: Zoommed-in image for the same models. In the inset we further zoom in on the very final evolution phase, post-silicon core burning; for the MESA models markers indicte the central conditions  at times of 1000\,s (circles), 100\,s (squares), 10\,s (crosses), and  1\,s (stars) before collapse.}
    \label{fig:rhoT_all}
\end{figure*}

\section{Results}
\label{sec:results}

In the previous sections, we have presented the new extensions to the physics included in GENEC to allow the evolution of stellar models in the high density and high-temperature regimes typically found in the last stages of evolution of massive stars. We have thus employed this  new version  to produce a small sample of models that are run to the pre-supernova stage. The choice of models studied are a 15 solar mass model at solar metallicity, 20 and 25 solar mass models, both run at Large Magellanic Cloud (LMC) metallicity. To compare the results with the current state of the art, we also run the same models with MESA \citep{Paxton_Bildsten_Dotter_Herwig_Lesaffre_Timmes_2011, Paxton_Cantiello_Arras_Bildsten_Brown_Dotter_Mankovich_Montgomery_Stello_Timmes_et_al._2013, Paxton_2018, Paxton_Smolec_Schwab_Gautschy_Bildsten_Cantiello_Dotter_Farmer_Goldberg_Jermyn_et_al._2019}  using version 10398. The physics of the GENEC models not explicitly mentioned in this paper are the same as in \cite{Ekstrom_Georgy_Eggenberger_Meynet_Mowlavi_Wyttenbach_Granada_Decressin_Hirschi_Frischknecht_et_al._2012}. We point out, for example, the mass loss prescription and the mixing as two major physical ingredients that are unchanged. For the MESA models, we use convective boundary mixing (CBM) with a value of $f = 0.02$ at the top of the convective layer and $f=0.004$ below. This induces slightly more mixing than the prescription of GENEC. Thus, the MESA models are expected to behave as slightly more massive stars than the GENEC models with the same initial mass. The Schwarzschild criterion defines the convective boundaries in our model, and as such, we did not need to implement semi-convective mixing. The MESA models are run until the pre-SN link, defined as the point where the maximum infall velocity reaches $10^8\,\text{cm\,s}^{-1}$. For the GENEC models the acceleration term required to track the in-fall velocity is not implemented yet and, thereby, we do not share a similar definition of the stopping point. To maintain a consistent stopping point, however, we use the central temperature of the MESA model one second before core collapse, and take the GENEC model that first reaches  the same core temperature and define this as our pre-supernova model. This ensures that we do not evolve the GENEC models further than they should go due to the lack of core contraction (supplied by the acceleration term), and also that the defined pre-SN GENEC model is extremely close to core collapse. Hence, this model  can be followed up by hydrodynamic codes with ease.

\subsection{Core evolution}
\label{sec:core_evolution}

In this section, we focus on the evolution of central conditions for all models, as most of the novel improvements in GENEC affect the high-temperature and high-density conditions generally reached in the inner core.

\begin{figure}[htb]
    \centering
    \includegraphics[width=0.5\textwidth]{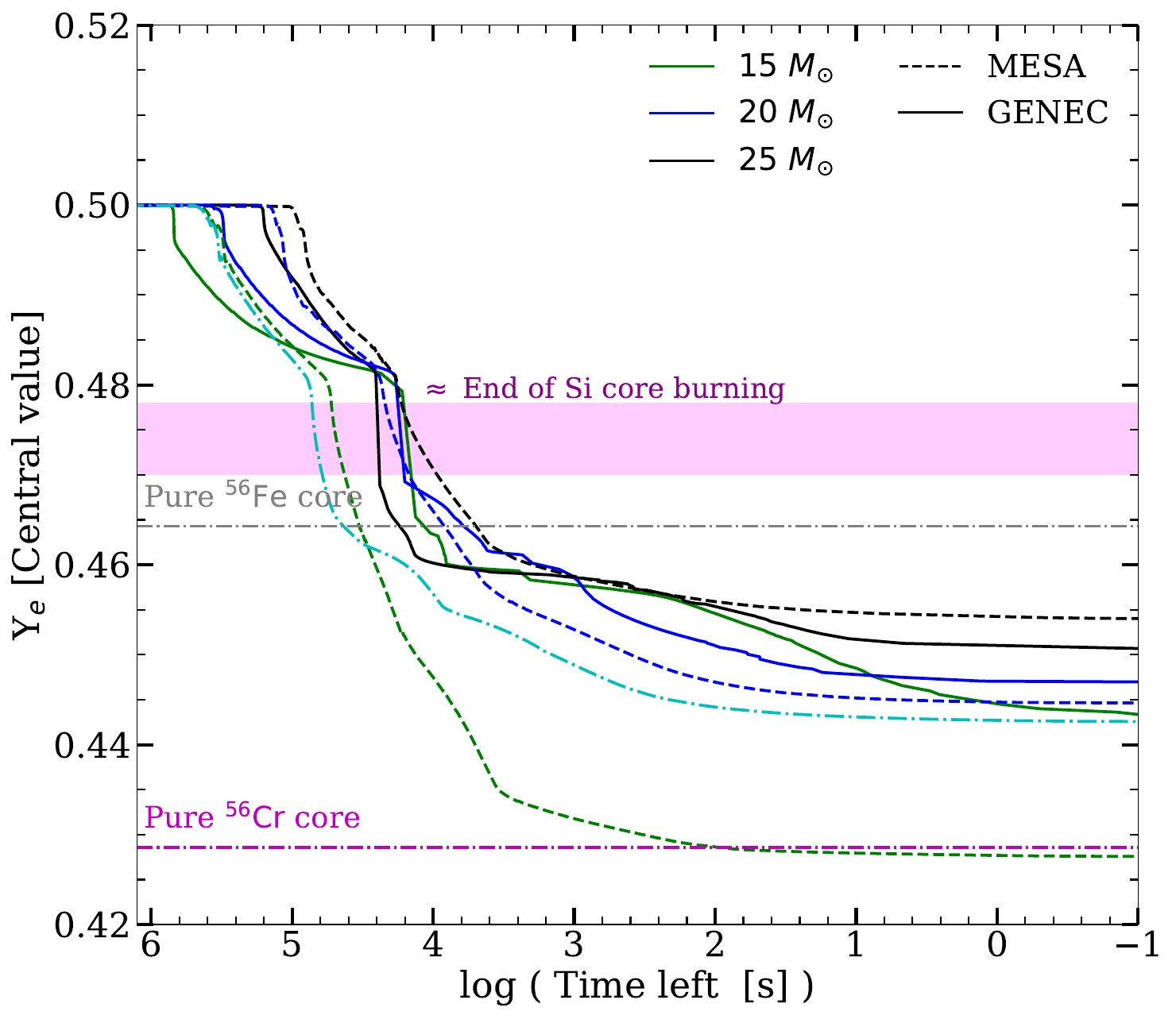}
    \caption{Evolution of the central value of the electron fraction for all models. The cyan model displays the $Y_e$ evolution of a $15M_\odot$ MESA model with ten times slower electron capture rates. The horizontal dot-dashed lines show the value of $Y_e$ that the centre of the star would attain if it were composed of pure $\rm^{56}Fe$ (grey) and $\rm^{56}Cr$ (magenta). The $\rm^{28}Si$ core burning begins, approximately, when $Y_e$ decreases from 0.5. We indicate the interval of the end of Si core burning by the magenta region.}%
    \label{fig:Ye_evol}%
\end{figure}

In Fig.~\ref{fig:rhoT_all} we compare the evolution of central temperature versus central density for our different models. In the left panel, we separate by the red dotted lines the region where the D-EoS is applied and where the T-EoS is applied, as defined by Eq.~\eqref{eqn:EOS_cond}. We note that the transition across the boundary between the two segments of the EoS in Fig.~\ref{fig:rhoT_all} is free of numerical artifacts, reflecting the smoothness of the connection between the D-EoS and T-EoS under the prescribed matching conditions, we detail further  the matching point in Appendix~\ref{sec:Matching_EOS}. In the right panel, we zoom in on the later stages of evolution, where the differences between models are largest. The general aspect of the tracks looks qualitatively (and quantitatively) like the ones shown in Fig.~1 of \cite{Heger_2001PhRvL..86.1678}, computed with the KEPLER stellar evolution code. Comparing the GENEC models to the MESA ones, here we see that most of the same aspects are present in each track, like the various loops and bends occurring at the start of consecutive burning stages in the core. However, the size of the loops and position of bends are not exactly the same from code to code. These differences generally stem from slightly different treatments of convection in the core and in the layers above it.

In the GENEC models, the convective core size is determined firstly by using the Schwarzschild criterion. Then core overshooting is applied only for the hydrogen and helium burning phases where the size of the convective core is increased by $\alpha H_p$, with $H_p$ being the pressure scale height and $\alpha=0.1$  as in \cite{Ekstrom_Georgy_Eggenberger_Meynet_Mowlavi_Wyttenbach_Granada_Decressin_Hirschi_Frischknecht_et_al._2012}. We note that this calibration was made to reproduce the Main Sequence observations of stars in the mass range $1.35M_{\odot} < M < 9 M_{\odot}$, and could require further fine-tuning for the stars considered here. We choose the overshooting and convection criterion in our MESA models to give similar behaviours for the treatment of convection in the initial burning phases. In the advanced burning stages, overshooting and undershooting is applied in the MESA models, which will likely contribute to differences between the codes. However, this choice was made to remain consistent with the MESA state-of-the-art. 

Locally, the position in the $(\rho_{\rm c}, T_{\rm c})$ plane can differ by $\sim 35\%$ among equivalent GENEC and MESA models. The overall core dynamics, marked by the ascension along a path where $\rho_{\rm c} \propto T_{\rm c}^{3}$, brings all models to relatively similar (though not equal) pre-supernova conditions.  During the very final approach to collapse, the last straight path in the profiles before reaching $T\approx10^{9.9}$\,K shows a very similar slope. In the inset of Fig.~\ref{fig:rhoT_all} we show the very final contraction of the stellar core before collapse. The evolution of MESA models and GENEC models share the same slope, but with slightly different central conditions, pointing towards a mostly adiabatic contraction detached from the conditions of the rest of the star. In Fig.~\ref{fig:rhoT_all} we place markers for the MESA models indicating time before collapse of 1000s (circle), 100s (square), 10s (cross) and finally 1s (star). Along this final branch, the changes in the core are extremely rapid. The final model of GENEC is chosen to match the central temperature of the MESA model that is one second away from collapse, i.e. marked by the star in Fig.~\ref{fig:rhoT_all}.
At the very tip of these evolution tracks, we see a slightly faster contraction for MESA models, which is indicative of the acceleration effects becoming dominant and the collapsing of the core. This is shown clearly in Fig.~\ref{fig:rhoT_all} by the magenta dashed-dotted curve, which is the same $15M_{\odot}$ mass model run in MESA but without using the acceleration term in the stellar evolution equations. The curve of this model and of the other $15M_{\odot}$ MESA model are identical for most of the evolution — we only display the very end phase for this reason — excepted the last second before collapse. Past the star marker we see the model with the acceleration term contracts slightly faster (i.e. the central density rises faster for a given temperature increase), whereas the model without the acceleration term stays on an adiabatic contraction. This indicates that the absence of the acceleration term in GENEC does not affect the evolution until approximately the last second before collapse, at which point a transition towards fully coupled hydro-codes is preferable to follow the collapse anyway.

The electron fraction evolution of the core leading up to collapse is shown in Fig.~\ref{fig:Ye_evol}. The effect of electron capture in our reduced network only becomes relevant once core $\rm^{28}Si$ burning begins. The corresponding decrease in the electron fraction of the core is followed in our simplified  GeValNet25 network by the electron capture chain, detailed in Sect.~\ref{sec:Network}. A large drop in $Y_e$ generally occurs by the end of $\rm ^{28}Si$ core burning, where the inner core is entirely made up of iron group elements and as such susceptible to electron capture. In Fig.~\ref{fig:Ye_evol} we highlight this region by a magenta band showing roughly the depletion of $\rm^{28}Si$ in the core. The electron fraction evolution flattens as the model approaches collapse, as expected due to the quasi beta equilibrium that is reached in the core \citep{Heger_2001ApJ...560..307}. The final value of $Y_e$ is heavily dependent on the mixture of $\rm^{56}Fe$ and $\rm^{56}Cr$. Therefore, the precise value of the central $Y_e$ obtained in these models is closely related to the rates used for the modelled reaction $\rm^{56}Fe \ \xrightarrow{(2)}\ \rm ^{56}Cr$. This chain is only a rough approximation for how the electron fraction changes due to weak reactions. In this work we intend only to reproduce the basic functionality of the reduced MESA approx21\_plus\_co56 network.
The current network aims only to reproduce the basic properties found in the pre-collapse phase of massive stars due to neutronisation. The final $Y_e$ reached by the GENEC models in the core differs from that of MESA, but are all within the range of larger and more sophisticated networks, \citep[see e.g.][]{Farmer_2016, Heger_2001PhRvL..86.1678}. The evolution of $Y_e$ and the values reached before core collapse are very similar for GENEC and MESA for models with masses $\ge 20\,M_\odot$, while there is some moderate discrepancy for the smaller mass under consideration ($15\,M_\odot$). To further understand this discrepancy, the cyan dash-dotted line in Fig.~\ref{fig:Ye_evol} shows the same $15M_{\odot}$ MESA model but run with electron captures that is ten times slower between $\rm^{56}Fe$ and $\rm ^{56}Cr$. We recall here that in reduced-type networks, the speed of the final part of the electron capture chain is calibrated by an artificial factor. In the reduced MESA approx21\_plus\_co56 network, this is by default $10^{-4}$. The extra model shown in Fig.~\ref{fig:Ye_evol} is run with $10^{-5}$, which is the same artificial factor that the GENEC models are run with. We see that slowing down the final step of the electron capture chain, the $Y_e$ evolution becomes much more similar to that of the $15M_{\odot}$ GENEC. This demonstrates some sensitivity of the central value, and thus the core structure, of $Y_e$ to this factor. Furthermore, using a single value for this calibration factor over the entire range of massive stars is most likely not adequate to imitate the real behaviour of larger electron capture networks. It is clear that in both reduced nuclear networks (GENEC and MESA) the treatment of electron captures is approximate, and these captures tend to be most important in the low-mass end of the massive star distribution. Therefore, approximating with the same recipe over all masses may lead to discrepancies. A more complete electron capture network (optimally including also beta decays) is required to accurately track the core neutronisation.

\begin{figure}[ht]
    \centering
    \includegraphics[width=0.48\textwidth]{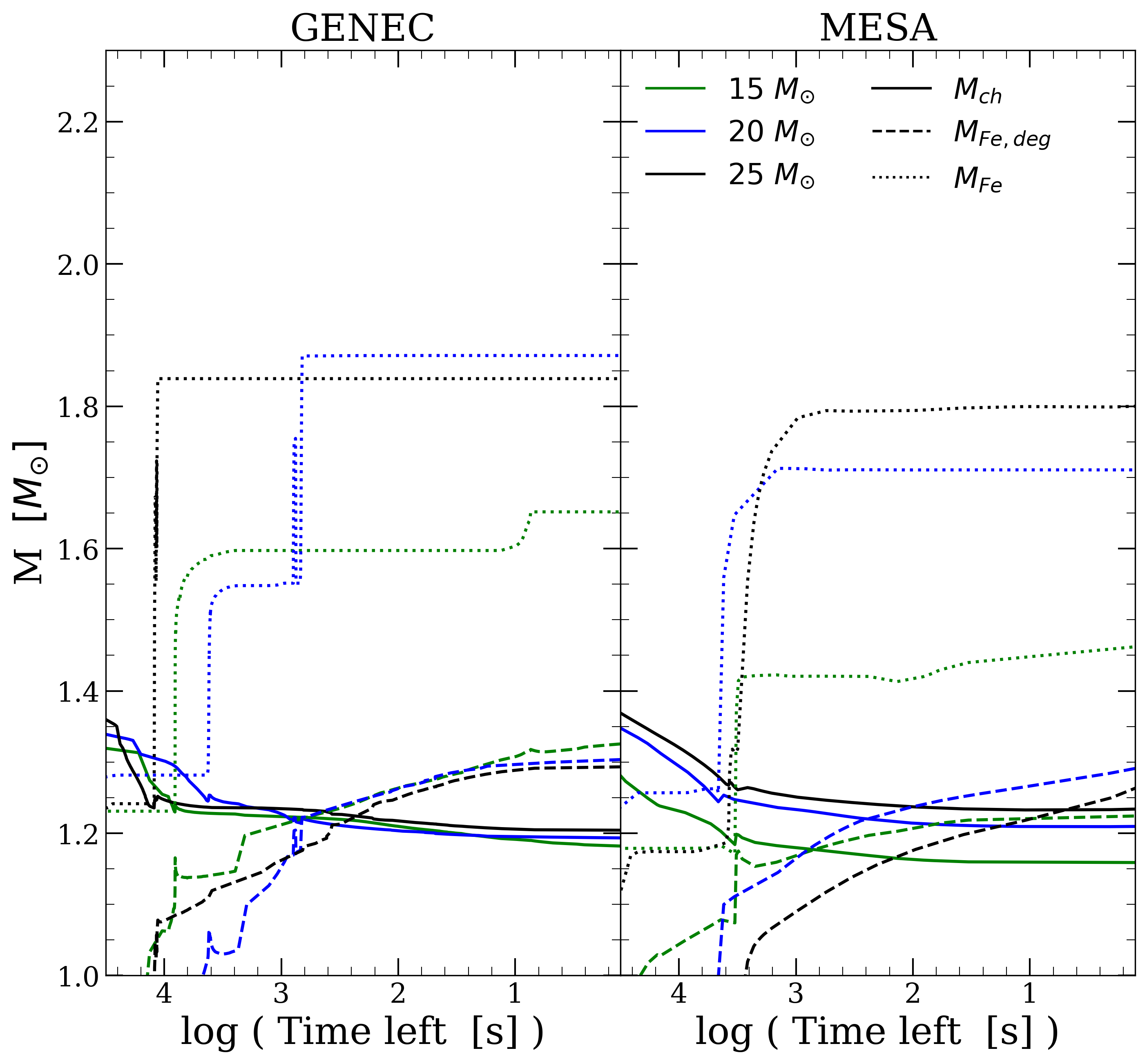}
    \caption{Evolution for all models of the Chandrasekhar mass ( solid lines), computed using Eq.~\ref{eqn:M_ch}. Also shown is the iron core mass (dotted line) defined by the mass coordinate where the sum of elements heavier than $\rm^{48}Cr$ goes below 0.5. Furthermore, we show the degenerate iron core mass that we defined in Eq.~\ref{eqn:M_fe_degen}. In the left panel are shown the GENEC models and in the right the MESA models.
    }
    \label{fig:chandra_evol}%
\end{figure}

\begin{figure*}[ht]
    \centering

\includegraphics[width=0.9\textwidth]{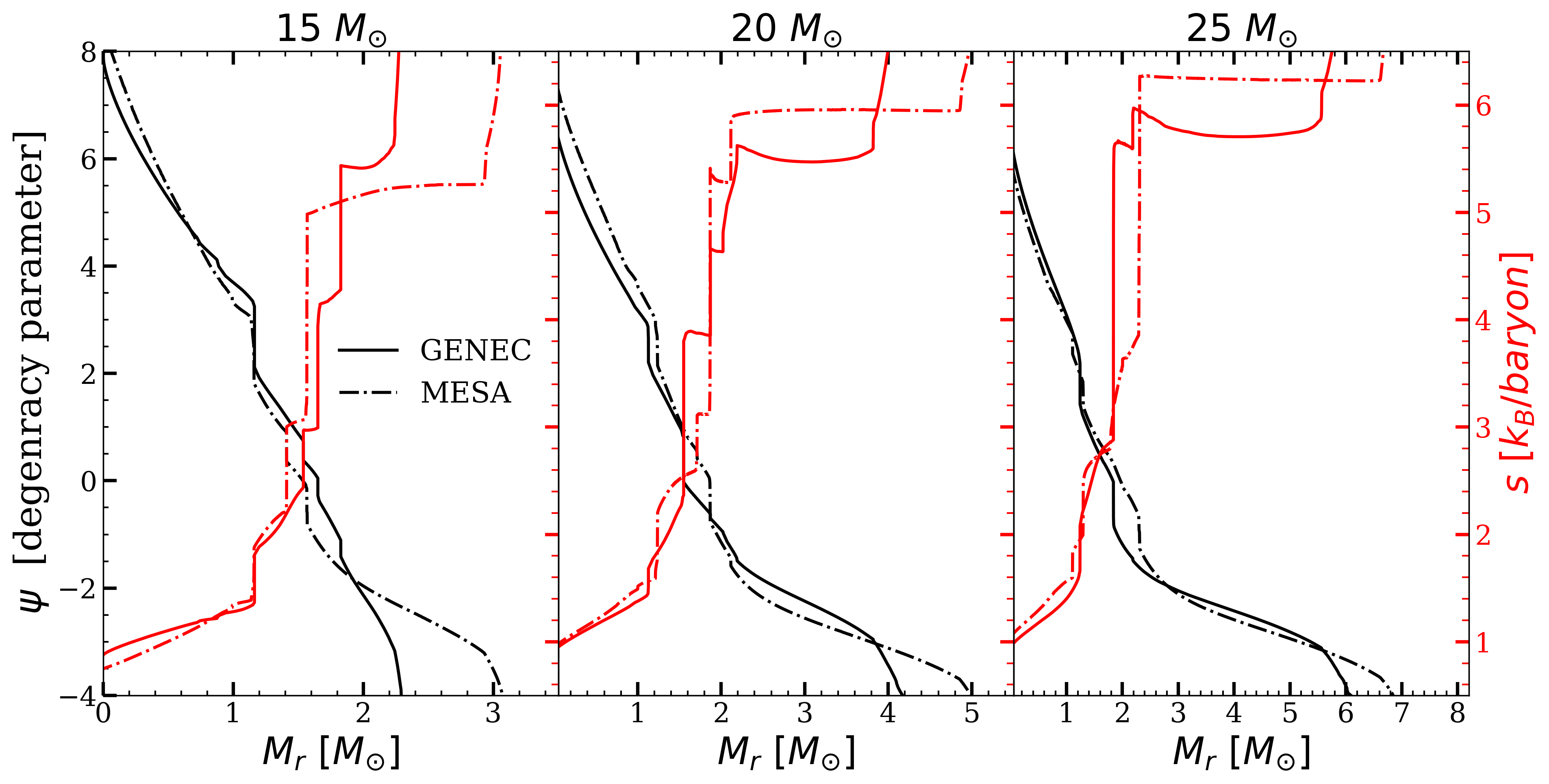}
   \caption{Entropy per baryon (red) and the degeneracy parameter $\Psi$ (black), as a function of mass for the pre-collapse models (see Appendix~\ref{sec:Dichte_EOS} for details on the degeneracy parameter). The solid and  dash-dotted lines correspond to the GENEC and MESA models, respectively. From left to right, we show the $15M_{\odot},  20M_{\odot}$, and $25M_{\odot}$ models.}
    \label{fig:psi_entropy}
\end{figure*}

The decrease in the electron fraction of the stellar core leads to a decrease in the electronic contribution to the pressure. At these late stages, the iron core is strongly degenerate, and the electron pressure is maintaining the integrity of the core. As shown by numerous theoretical studies, originating from \cite{Chandrasekhar_1939}, approximately when the iron core reaches the Chandrasekhar mass, collapse will begin. At first order this critical mass is
\begin{align}
\label{eqn:M_ch0}
    M_{\rm ch,0} = 1.44 \left(\frac{Y_e}{0.5}\right)^2.
\end{align}In \cite{Baron_Cooperstein_1990ApJ...353..597} the authors investigate the role of this critical mass and add different corrections to the simplified estimation of Eq.~\ref{eqn:M_ch0} written as
\begin{align}
\label{eqn:M_ch}
    M_{\rm ch} = M_{\rm ch,0} \left[1 - 0.057 + \left(\frac{S_{e}}{\pi Y_e}\right)^2 +1.21\frac{1}{\rm A}S_e \right],
\end{align}
where $S_e$ is the entropy of electrons per nucleon and $\rm A$ the mass fraction weighted atomic number. 
For the iron core,  we define it in the same manner as is done in the massive star evolution literature, \citep[see e.g.][]{Hirschi_2004,Chieffi_Limongi_Straniero_1998, Paxton_Bildsten_Dotter_Herwig_Lesaffre_Timmes_2011,Heger_2001ApJ...560..307}. The edge of the iron core is taken where the sum of mass fractions of all elements heavier than $^{48}\rm Cr$ goes below 0.5, and we denote the enclosed mass by $M_{\rm Fe}$. However, for the purposes of comparing this mass with the Chandrasekhar mass, we also consider the degeneracy of the matter contained within the iron core. The original Chandrasekhar computation is done in the context of fully degenerate matter, which is not necessarily the case for what most authors define as the iron core. Therefore, we compute here $M_{\rm Fe,deg}$, defined by 
\begin{align}
\label{eqn:M_fe_degen}
    M_{\rm Fe,deg} = \int_0^{M_{\rm Fe}} f(\Psi)dm,
\end{align}
where $f(\Psi)$ is a weighting function varying from 0 to 1 with the expression,  
\begin{align}
    f(\Psi) = 1 - \frac{1}{1+\exp(\Psi - \frac{\Psi_0+\Psi_1}{2}) },
\end{align}
where $\Psi_0 = -4$ and $ \Psi_1=7$.\footnote{These values approximately delimit fully degenerate and non-degenerate matter. The Appendix.~\ref{sec:Dichte_EOS} gives further detail on this point.} 
In Fig.~\ref{fig:chandra_evol} we show the evolution of the Chandrasekhar mass, the iron core mass and the degenerate iron core mass in the last hours of evolution of our models. We observe that, for all masses, the iron core mass goes above the Chandrasekhar mass some $10^4$\,s before collapse for GENEC models and roughly $10^{3.5}$\,s before collapse for MESA models. The iron core becomes much heavier than the  Chandrasekhar mass when the first silicon shell burning produces enough iron elements so that our definition then includes this shell as part of the iron core. However, the matter in this shell is not necessarily degenerate and should not be fully counted as such when comparing with the Chandrasekhar mass. Comparison of $M_{\rm ch}$ with $M_{\rm Fe,deg}$ is more relevant to understand the collapse dynamics. We see that eventually the degenerate mass as well becomes larger than the Chandrasekhar mass, some 10's or 100's of seconds before collapse. There are strong variations in the iron core mass between codes here, as has also been noted before in \cite{Paxton_Bildsten_Dotter_Herwig_Lesaffre_Timmes_2011}, but a much tighter spread in terms of the degenerate core mass.  Qualitatively, the GENEC and MESA models behave in the same way, i.e. the 15\,$M_{\odot}$ model is the first to have $M_{\rm Fe,deg}$ larger than $M_{\rm ch}$ and the 25\,$M_{\odot}$ model is the last. This is due to matter being more degenerate in the core for models with smaller starting mass. All models reach masses of $M_{\rm Fe,deg}$ higher than the $M_{\rm ch}$ at similar times before the last model, showing that iron core collapse is imminent for GENEC and MESA progenitors alike.

\subsection{Pre-collapse structure}
\label{sec:precol}

\begin{figure*}[ht]
    \centering

\includegraphics[width=0.9\textwidth]{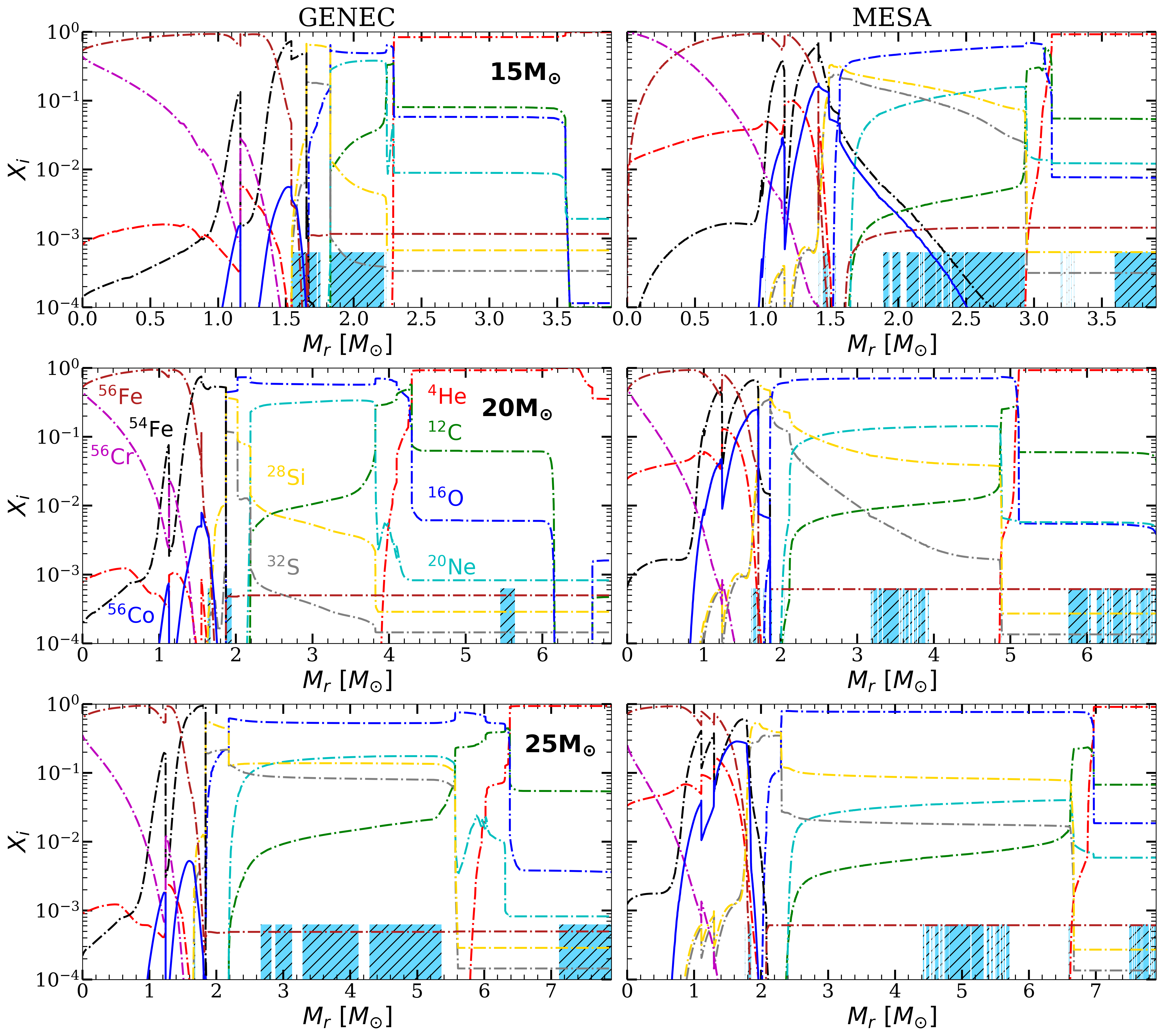}
   \caption{Chemical structure for the pre-supernova models. Shown are the species $^4\rm He$,$^{12}\rm C$,$^{16}\rm O$,$^{20}\rm Ne$,$^{28}\rm Si$,$^{32}\rm S$,$^{54}\rm Fe$,$^{56}\rm Fe$,$^{56}\rm Co$ and $^{56}\rm Cr$ as a function of enclosed mass. We also highlight the convective regions with the light blue hashed regions. \\}
    \label{fig:Model_structure}
\end{figure*}

\def\arraystretch{1.2}%
\begin{table*}[!ht]
\centering
\caption{Properties of pre-supernova models. 
}
  \begingroup
  \setlength{\tabcolsep}{5pt} % Default value: 6pt

    \begin{tabular}{ c c c c c c c c c }

        \hline 
        \hline
        Models & \multicolumn{3}{c}{$15 M_{\odot}$} & \multicolumn{3}{c}{$20 M_{\odot}$} &  \multicolumn{2}{c}{$25 M_{\odot}$} \\\cline{2-3}\cline{5-6}\cline{8-9}

        & GENEC & MESA & & GENEC & MESA & & GENEC & MESA \\

        $M_{\rm tot} \ [M_{\odot}]$ & 13.29 & 11.47 & &18.50& 17.32 & &17.31 & 19.95\\
        $\log(R_{\rm tot} / R_{\odot})$ & 2.80 & 3.02 &&2.91& 3.08 & & 3.00 &  3.15\\
        $\log(L / L_{\odot})$ & 4.88 & 5.02 && 5.21& 5.28 &  & 5.41 & 5.43\\
        $\log(T_{\rm eff} [{\rm K}] )$ & 3.55 & 3.51  &&3.58& 3.54& & 3.61 & 3.56\\
\hline 
        $\rm M_{\rm He,shell}\ [M_{\odot}]$ &4.12 & 4.87 &  & 6.65 & 7.20 & &  8.85  & 9.39\\
        $\rm M_{\rm CO,shell}\ [M_{\odot}]$ & 2.29 & 3.12 & &  4.30 & 5.10 && 6.38 & 6.97 \\
        $\rm M_{\rm Si,shell}\ [M_{\odot}]$ & 1.83 & 1.60 & & NA & 1.88 && 2.19 & 2.30  \\
        $\rm M_{\rm Fe,core}\ [M_{\odot}]$& 1.65 & 1.49 & & 1.87 & 1.71 & & 1.84  & 1.82\\
\hline
        $f_{\rm Si,O}$ & 0.0848 & 0.400 & & 0.379 & 0.338 & & 0.247 & 0.0493\\
\hline
        $Y_{e,\rm c}$ & 0.446 & 0.427 & & 0.447 & 0.444 & & 0.451 & 0.453\\
        $Y_{e,\rm core}$ & 0.463 & 0.458 & & 0.467 & 0.467 && 0.466 & 0.470\\
        $M_{\rm Ch} \ [M_{\odot}]$ & 1.18 &1.15 & & 1.19 & 1.22 & & 1.22 & 1.24 \\
        $M_{\rm Fe,deg} \ [M_{\odot}]$ & 1.31 &1.31 & & 1.30 & 1.39 & &  1.30 & 1.39 \\
        $\Psi_c$ & 7.92 & 10.4 & & 6.86 & 7.87 & & 6.53 &6.87 \\
        $s_{\rm c} \ [\rm kb/baryon]$ & 0.872 & 0.749 & & 0.907 & 0.936 & & 0.943 & 1.05\\
        $\log(\rho_{\rm c}\ [\text{g\,cm}^{-3}  ])$ & 9.65 & 10.2 & & 9.42 & 9.78 && 9.36 &9.59\\
        $\log(T_{\rm c} \ [{\rm K}])$ & 9.92 & 10.0 & & 9.90 & 9.97 && 9.90 & 9.96\\
\hline 
        $\rm \xi_{Mr=2.5}$ & 0.0375 & 0.128 & & 0.267 &  0.240 & & 0.289 & 0.422\\
        $\rm \xi_{s=4}$ & 0.671 & 0.712 & & 0.493 & 0.750 && 0.651 &  0.589 \\
        $M_4 \ [M_{\odot}]$ & 1.67 & 1.57 & & 1.87 & 1.87 && 1.84 & 2.29\\
        $\mu_4$ & 0.0126 & 0.0888 & &0.158 & 0.129 && 0.154 & 0.130\\
\hline
     $M_{4,g}$ & 1.49  &1.41   & & 1.65 &  1.62 & &  1.65 & 1.97 \\
\hline
\hline\\
    \end{tabular}
    \endgroup
\tablefoot{
 $M_{\rm tot}$ is the total mass of the star, $R_{\rm tot}$ the total radius, $L$ the luminosity at the surface and $T_{\rm eff}$ the surface effective temperature. The shell masses correspond to the outer mass coordinate where the element in question is most abundant; NA is used if at no point in the star is that species the most abundant. $f_{Si,O}$ is as defined in Eq.~\eqref{eqn:fsi_o}. The thermodynamic variables $\rho$,   $s$, $Y_e$, and $T$ evaluated at the innermost point of the star are annotated with subscript "c". Quantities with subscript "core" correspond to mass averaged quantities over the iron core, e.g. $Y_{e,\rm core} = \frac{1}{M_{\rm Fe,core}}\int_0^{\rm M_{\rm Fe,core}} Y_e(m) \ dm$. $M_{\rm Ch}$ is the Chandrasekhar mass as defined in Eq.~\eqref{eqn:M_ch}. The explodability parameters $\xi_M$, $M_4$ and $\mu_4$ are defined in Eqs.~\eqref{eqn:compactness},~\eqref{eqn:M4} and \eqref{eqn:mu4}, respectively. $M_{4,g}$ is the gravitational mass of a hypothetical neutron star resulting from the collapse of the core (Eq.~\eqref{eq:M4g}).}

\label{table:pre_sn_compar}
\end{table*}

As we have seen in the previous section, the very final hours of evolution in both GENEC and MESA models will lead to a core collapse due to neutronisation of the core and reaching the limits of instability with respect to the Chandrasekhar mass of the core. In its current version, GENEC does not include the acceleration term in the mechanical equilibrium equation. Without this term, we cannot define a pre-SN model stopping point (the pre-SN link), as is commonly done in the literature, when the maximum in-fall velocity reaches 1000\,$\rm km \, s^{-1}$. As explained in the previous sections,  we take as pre-SN link definition for the GENEC models the profile where the central temperature is the same as the MESA model equivalent one second before reaching the 1000\,$\rm km \, s^{-1}$ infall velocity limit. This is the model shown by the star in the inset of Fig.~\ref{fig:rhoT_all}. In this section, we compare the structure of the final models produced using GENEC and MESA and highlight the interesting properties displayed for supernova progenitors. A summary of the characteristics discussed in this section is provided in Table~\ref{table:pre_sn_compar}. 

In Fig.~\ref{fig:psi_entropy} we show the profiles of entropy and the degeneracy parameter $\Psi$ for the final models. The entropy within the inner 1\,$M_{\odot}$ is almost identical between codes, large differences appear in the outer regions where the exact boundaries between the burning shells, and the strength of mixing, plays an important role. The general trend of degeneracy is consistent between codes, and goes as expected, i.e. the 15\,$M_{\odot}$  being the most degenerate and the 25\,$M_{\odot}$ the least. The 15\,$M_{\odot}$ model of MESA is more degenerate in the core, which can also be seen in Fig.~\ref{fig:rhoT_all} where the MESA model reaches higher densities for a given temperature.
We note that the shell structure beyond the inner core of roughly $1\, M_{\odot}$, differs between codes. For example, there are two steps in the entropy for the GENEC $25\,M_{\odot}$ at $M_r = 2 M_{\odot}$ and then at $M_r = 2.5 M_{\odot}$,  whereas the MESA model only shows one step. Similar differences can be observed in the other models. These stem from different evolution of the various shell burning events. They can be due to variations in the nuclear reaction network, which use different rates and compute different energy generation values, or also differences in the treatment of convection and overshooting. 
The main differences between the models at collapse, between different codes, is not in the central values of physical quantities, but rather in the stellar structure which can have large consequences for the type of collapse and explosion that may ensue.

In Fig.~\ref{fig:Model_structure} we show the chemical composition for all pre-supernova models. There are numerous differences both between codes and between models of different masses. We point out the generally smaller C/O core for GENEC models (i.e. the helium shell begins at lower masses in GENEC than in MESA; see also Table.~\ref{table:pre_sn_compar}). This is caused by overshooting differences in the earlier phases. The $\rm ^{28}Si$ shell for the GENEC 20\,$M_{\odot}$ model disappears, with a part going to the iron core and the rest merging with the oxygen shell. This happens in no other models; however, the merging of the Si/O shell has begun in the MESA 15\,$M_{\odot}$ where we see silicon being mixed out until $M_r\approx 3 M_{\odot}$. With the simplistic network used here, the inner core composition is dominated by $\rm ^{56}Fe$ and $\rm ^{56}Cr$. The relatives abundances of these two species mostly shapes the $Y_e$ profile in the core. We may point out here that the MESA models show a higher helium abundance in the core than the GENEC models. This is due to the GeValNet25 implementation not including the necessary reactions to link the QSE group of neutrons, protons, and helium needed in the very final phase. This omission speeds up significantly the nuclear calculations and does not alter significantly the evolution of $Y_e$ (which is, anyway, only approximate in our code because of the effective treatment of the weak reactions).
Furthermore, we indicate by the light blue hashed regions where convection is active in the star in Fig.~\ref{fig:Model_structure}, the active convection regions vary significantly in all models and provide no clear rule of predicting at collapse what shells are unstable to convection. 

\begin{figure*}[ht]
    \centering
    \includegraphics[width=0.48\textwidth]{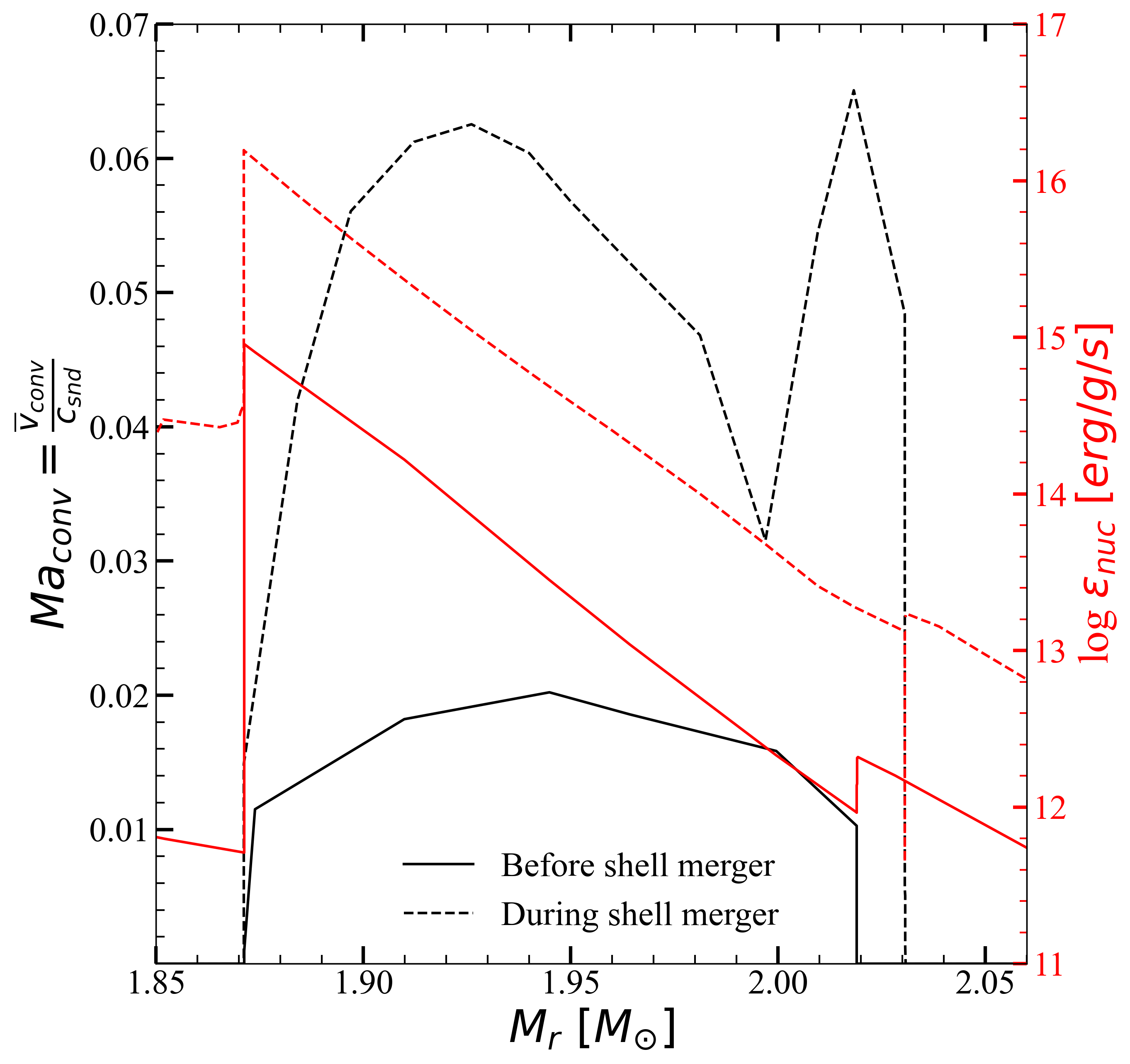}
    \includegraphics[width=0.48\textwidth]{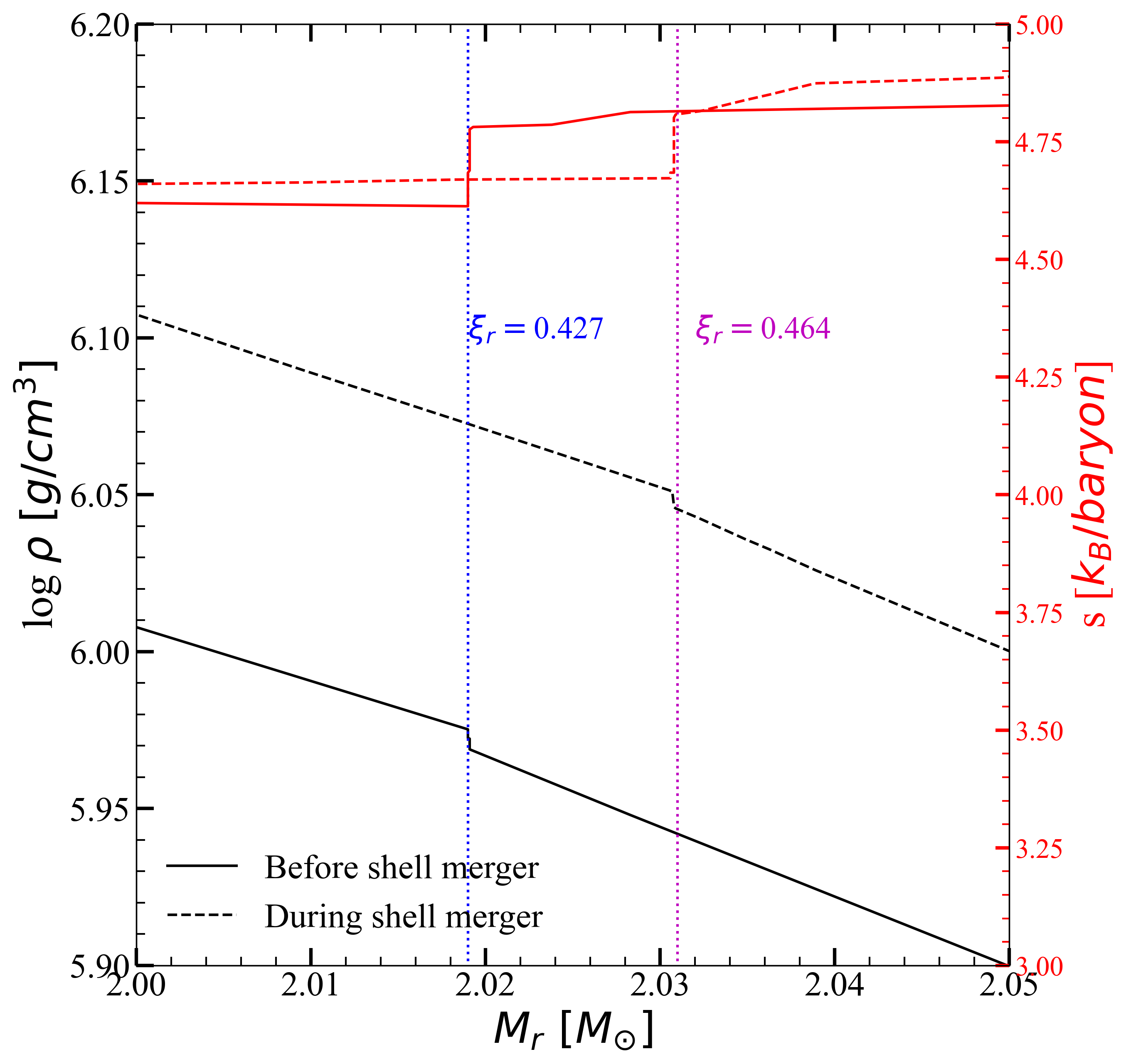}
    \caption{Properties of the 20\,$M_{\odot}$ GENEC model  before and then during a silicon-oxygen shell merger event occurring very shortly before collapse. The time gap between the two models shown is 15\,s.  On the left we show the convective Mach number, i.e. the ratio of the velocity predicted by mixing length theory to the sound speed, and also the energy produced by nuclear burning. On the right is shown the density and entropy structure of the shell in question.
    }
    \label{fig:Shell_merger}
\end{figure*}

Between codes, the structure of the silicon and oxygen shells at the point of collapse can vary quite significantly. As in \cite{Yoshida_Takiwaki_Kotake_Takahashi_Nakamura_Umeda_2019}, we define  the quantity $f_{\rm Si,O}$,
\begin{align}
\label{eqn:fsi_o}
f_{\rm Si,O} &= C_0 \int_1^{10} X_{^{28}\rm Si}X_{^{16}\rm O}  \nonumber \\ & 
\times \Theta(X_{^{16}\rm O}-0.1)\Theta( X_{^{28}\rm Si}-0.1)\rho r_8^2 dr_8,
\end{align}
where $\Theta(x)$  is the Heaviside function($\Theta(x)=1$ for $x\ge 0$; $\Theta(x)=0$ for $x< 0$) and $r_8$ the radius in units of $10^8$ cm. The constant $C_0$ is set as, 
\begin{align}
\label{eqn:c0}
C_0^{-1} &= \int_1^{10} X_{^{28}\rm Si}X_{^{16}\rm O} \rho r_8^2 dr_8,
\end{align}
so that $f_{\rm Si,O}=1$ if the whole region from $r_8=1$ to $r_8=10$ has $X_{^{16}\rm O}>0.1$ and $X_{^{28}\rm Si}>0.1$.
 This helps quantify the amount of potential turbulent mixing in the silicon-oxygen shell at the pre-supernova stage. When such mixing occurs as the collapse of the core begins can lead to impacting the  properties of the subsequent explosion. It also may indicate the need for a better follow up of the model in multi-dimension simulations prior to collapse (see Sect.~\ref{sec:discussion}).
The values for the pre-supernova models are given in Table.~\ref{table:pre_sn_compar}. We note that the higher the value of $f_{\rm Si,O}$, the larger the overlapping region  between oxygen and silicon shells. In practice, if these shells are overlapping, a merger event is occurring. Hence, a value of  $f_{\rm Si,O}$ close to 1 is indicative of the potential for strong turbulent mixing in the Si/O shell. The speed of this mixing can be very large, in Fig.~\ref{fig:Shell_merger} we show the value of the convective Mach number in such a shell reaching similar values as multidimensional studies such as \cite{2021_Varma_Muller} and \cite{2021_Yoshida} where the turbulent velocity in the Si/O rich layer was reaching several $100\,\rm km\,s^{-1}$ causing very rapid mixing with convective turnover timescales in these shells of order 10's of seconds.

\begin{figure*}[ht]
    \centering
    \includegraphics[width=0.48\textwidth]{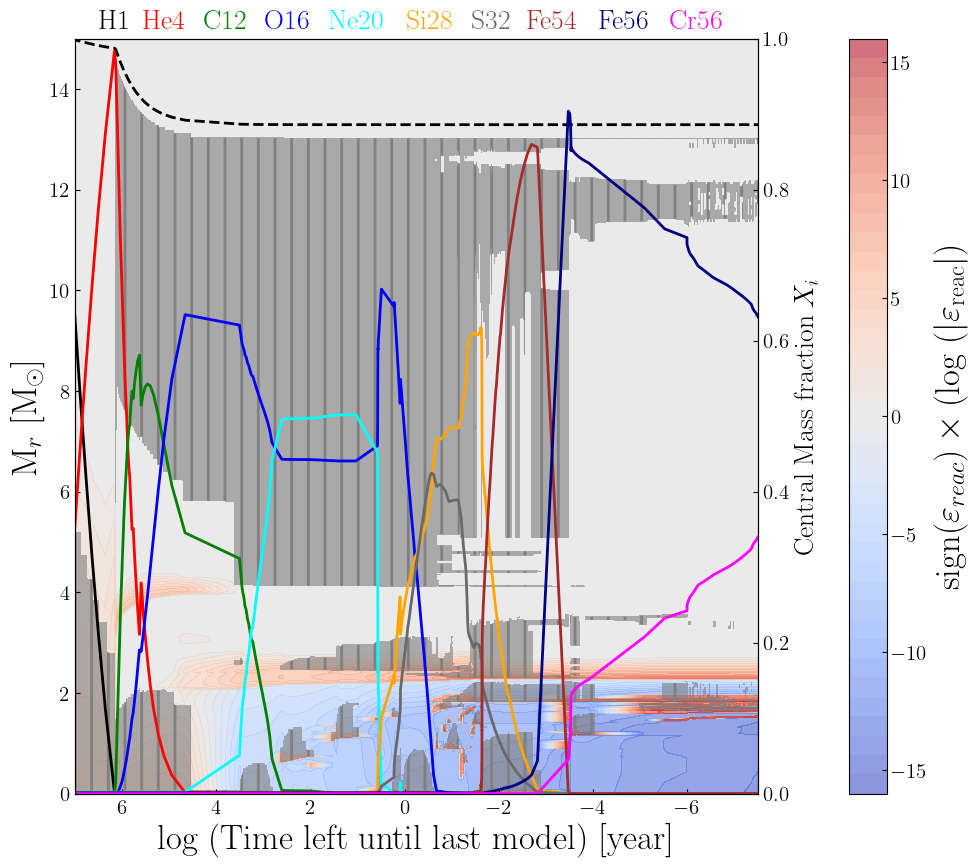}
    \includegraphics[width=0.48\textwidth]{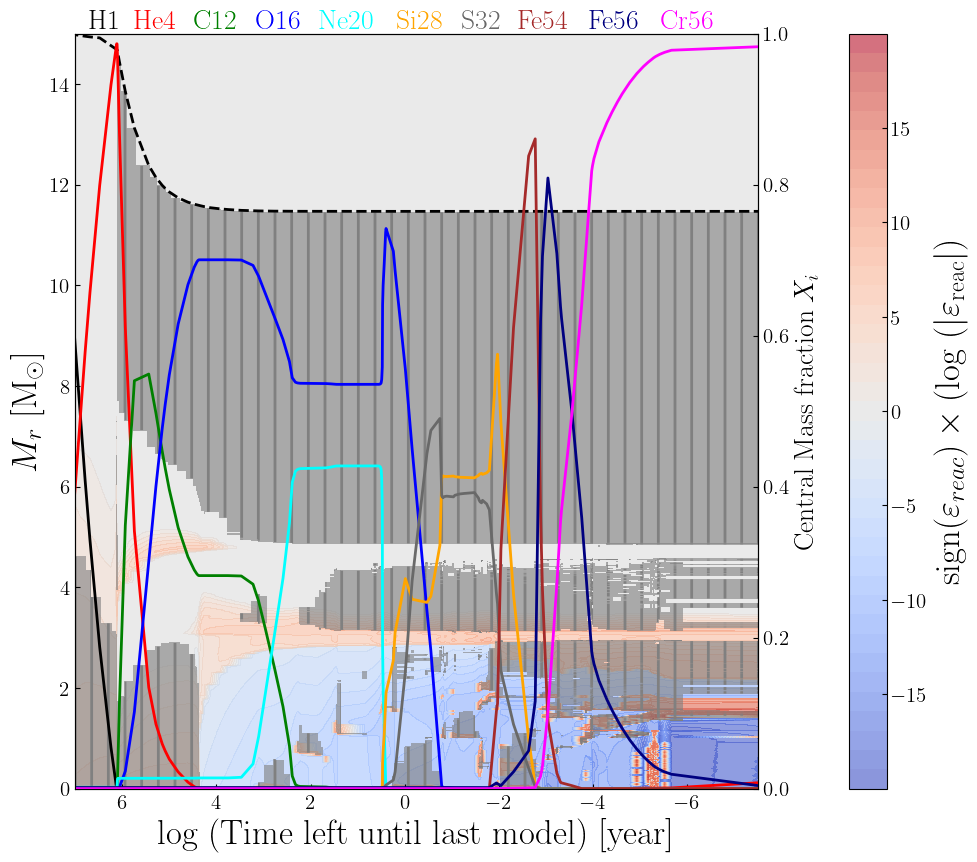}
    \caption{Space-time diagrams, or Kippenhahn diagrams, of the $15M_{\odot}$ model for GENEC (left) and MESA (right). The grey zones show convective regions. The plot background shade displays the total energy input, the nuclear energy subtracted by the loss of thermal neutrinos in log scale. Blue represents negative energy contributions and red positive contributions . On the right axis is shown the central abundance of a given selection of species to help identify the various burning phases. The black dashed line delimits the total mass of the model.}
    \label{fig:Kippenhan}
\end{figure*}

However, Si/O mixing is not necessarily happening for all models at the pre-collapse stage. For the 15$\,M_\odot$ model of GENEC there is a separated silicon shell from the oxygen shell, whereas the MESA 15$\,M_\odot$ is going through a merger of these two shells. This is also shown in the Kippenhahn diagram of Fig.~\ref{fig:Kippenhan}, where the MESA model has a large convective region from roughly 2 to 3 solar masses at $10^{-6}$\,yr before collapse.  

We know that shell mergers and enhanced mixing/burning prior to core collapse can have a strong impact on the stellar structure before collapse and thus impact the supernova explosion and evolution \citep{2016_muller,2017_jones,2022_rizzuti}. We can see this also in Fig.~\ref{fig:Shell_merger}. In this figure, we display the convective Mach number, entropy and density profiles of the 20\,$M_\odot$ GENEC model before and then during a silicon and oxygen shell merger event. These profiles are taken roughly $40\,$s before collapse, and there is a time difference of 15\,s between the two models shown. We see that during the shell merger there is a rapid amplification of the energy generation by nuclear burning, an increase by more than an order of magnitude in the entire shell. This is accompanied by a steep increase in the convective mixing as the convective Mach number increases by a factor 3 in the merged shell.
Furthermore, this merger event shifts the shell interface. The shift is only by $0.01 M_{\odot}$; however, we note that the compactness, $\xi_r=\frac{M_r[M_{\odot}]}{(r[1000\rm km] )}$, at the shell interface changes here by $8\%$ going from 0.427 to 0.464. These structural changes will impact the explosion dynamics. Therefore, knowing if the presence of these shell mergers before collapse are physical and are likely to happen is crucial. In this grid of models, shell mergers seem to be dependent on the stellar evolution code and the chosen prescriptions. These events are of course highly sensitive to differences in the network implementation and convective treatment, as such, exploration  in further detail on the trigger of these shell mergers and their impact on the explosion is required.

%Moved here to be in the order of reference in the text.

\subsection{Explodability conditions}

\begin{figure}[htb!]
    \centering
    \includegraphics[width=0.48\textwidth]{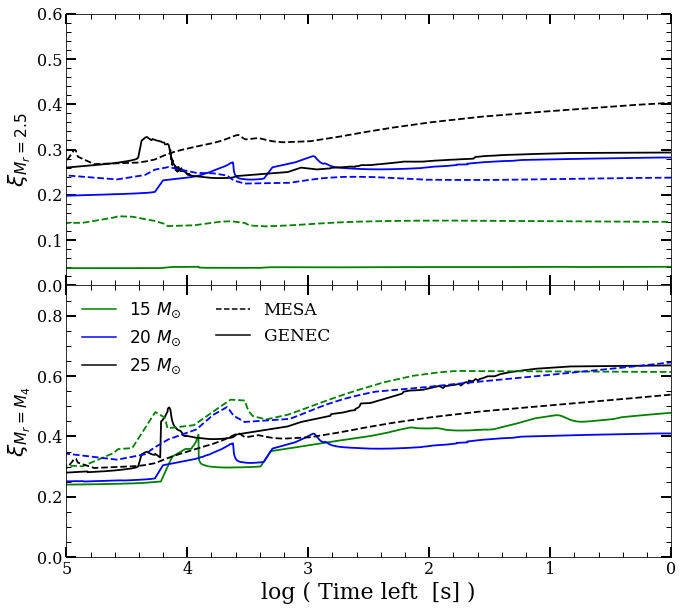}
    \caption{Evolution prior to collapse of different instances of the compactness parameter, $\xi_{Mr=2.5}$, $\xi_{M_r=M_4}$ (see Eq.~\eqref{eqn:compactness} for the definition.)}
    \label{fig:xi_evol}%
\end{figure}

In this section, we look at different criteria of explodability, which chiefly reflect the evolution of the innermost structure of each model, and help predict whether such models may undergo successful explosions or not. We define here a series of parameters often used to evaluate the explodability of 1D stellar evolution models. First the compactness parameter, $\xi_M$, often used as a basic criterion to estimate the fate of a star \citep{OConnor_Ott_2011ApJ...730...70},
\begin{equation}
\label{eqn:compactness}
    \xi_M \equiv \left.\frac{M/M_{\odot}}{R(M_{\rm b}=M)/1000\text{ km}}\right\vert_{t}
\end{equation}
where $R(M_{\rm b}=M)$ is the radial coordinate that encloses a baryonic mass at epoch $t$. The typical mass used to evaluate explodability is $M=2.5M_{\odot}$ and as close as possible to collapse.\footnote{\cite{OConnor_Ott_2011ApJ...730...70} suggested evaluating $\xi_{2.5}$ at the time of bounce, but \cite{Sukhbold_2014ApJ...783...10} observed that similar conclusions may be obtained evaluating it at the pre-supernova link and that the variation of the compactness after oxygen depletion is indeed relatively minor.} Following the work of \cite{Ertl_Janka_Woosley_Sukhbold_Ugliano_2016}  we also look at the behaviour of the parameters $M_4$ and $\mu_4$ defined as
\begin{equation}
\label{eqn:M4}
        M_4 \equiv M_r(s=4k_{\rm B})/M_{\odot},
\end{equation}
and
\begin{equation}
\label{eqn:mu4}
        \mu_4 \equiv \frac{\text{dm}/M_{\odot}}{\text{dr}/1000 \text{km}}\Bigg|_{s=4k_{\rm B}},
\end{equation}
where $M_r$ is the enclosed mass inside dimensionless entropy per nucleon of $s=4k_{\rm B}$. The \citeauthor{Ertl_Janka_Woosley_Sukhbold_Ugliano_2016} criterion consists of comparing $\mu_4$, which is a proxy for the mass accretion rate onto the compact remnant left after core-collapse, with the product $M_4\mu_4$, a magnitude correlated with the neutrino luminosity induced by mass accretion. This leads to a separation of models that produce neutron stars and models that produce Black Holes by finding which models successfully explode and which ones fail.

In Fig.~\ref{fig:xi_evol} we show the evolution of two instances of the compactness parameter, $\xi_{M_r=2.5}$, $\xi_{M_r=M_4}$, where $M_4$ is not a fixed mass position, but its value evolves in time and is defined as in Eq.~\ref{eqn:M4}. A general prediction is that the larger the compactness parameter $\xi_{2.5}$, the more likely it is that the progenitor collapses forming a black hole and does not produce a successful explosion. \cite{Ugliano_2012} concluded that black holes may result from stellar progenitors in which $\xi_{2.5} \gtrsim 0.35$, while models with $\xi_{2.5} \lesssim 0.15$  more likely produce neutron stars. For the range in between of these cases, $0.15\lesssim \xi_{2.5} \lesssim 0.35$, it is less certain when using only this one parameter. According to this criterion, the $25 M_{\odot}$ model will collapse to a black hole employing MESA, whereas the GENEC model is more within the range of uncertainty. It is noteworthy that $\xi_{2.5}$ is rising steadily prior to collapse, but reaches a lower value than  the MESA model. The $15 M_{\odot}$ mass model, for both codes, seems much more likely to produce a neutron star and finally the $20 M_{\odot}$ reaches similar values for both GENEC and MESA at the end of the evolution and lies within the uncertain range for this one parameter criterion. We notice that the compactness at the pre-SN link grows with mass, at least in the three cases at hand \citep[though there is no monotonic dependence of $\xi_{2.5}$ on ZAMS mass; c.f.][Fig.~4]{Ertl_Janka_Woosley_Sukhbold_Ugliano_2016}. GENEC models with $20\,M_\odot$ and $25\,M_\odot$ show  very similar values of $\xi_{2.5}$ close to collapse. 
To see the compactness evolution around the contracting core we also plot in Fig.~\ref{fig:xi_evol} the quantity $\xi_{M_r=M_4}$. The location where the entropy equals $4 k_{\rm B}$/baryon defines quite well the boundary of the iron core \citep[e.g.][]{Baron_Cooperstein_1990ApJ...353..597}. We see that this compactness rises for all models during the final hours of evolution, where the value of $\xi_{M_r=M_4}$ almost doubles for all models.

\begin{figure}[ht]
    \centering
    \includegraphics[width=0.5\textwidth]{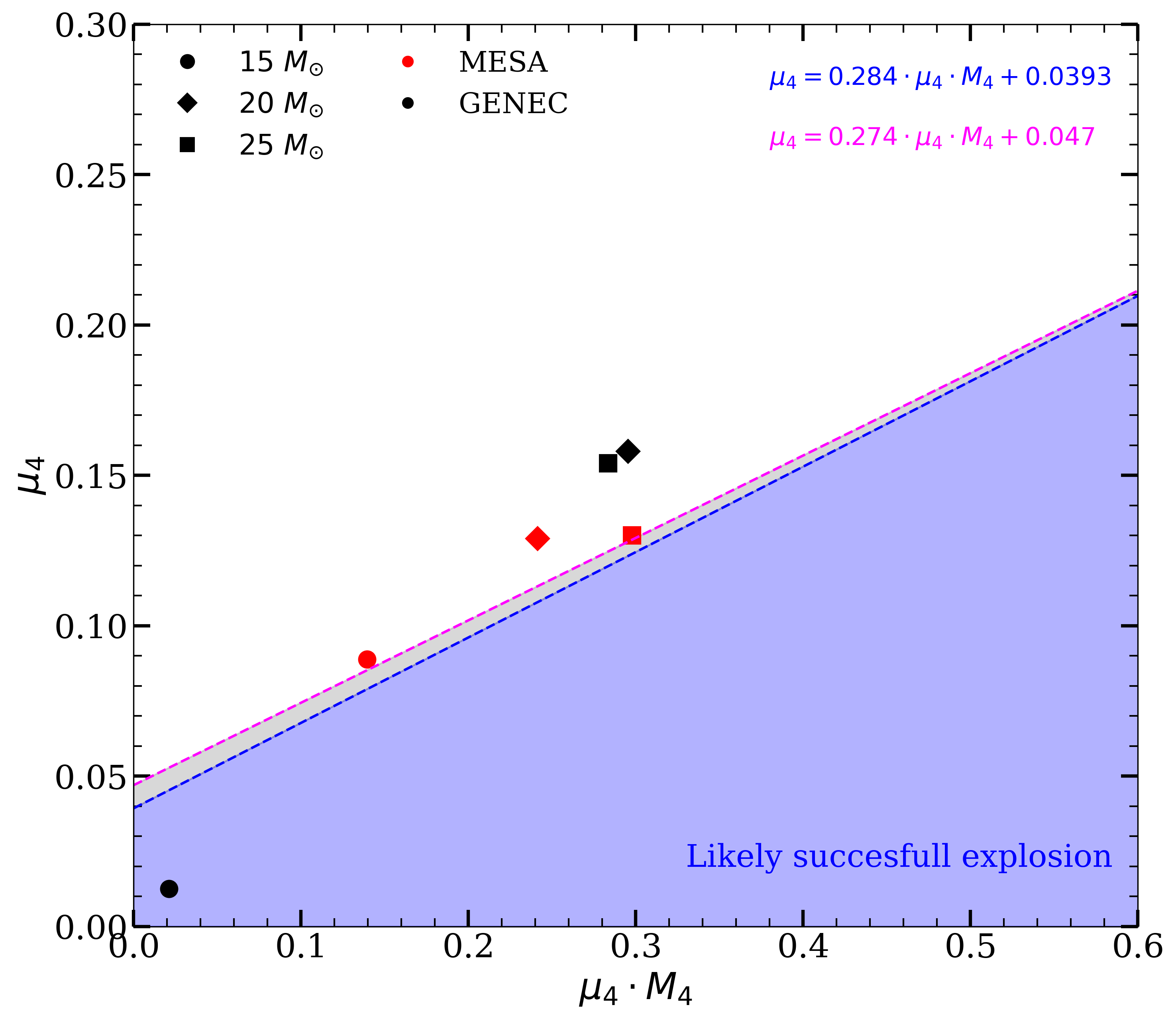}
    \caption{ Position for the pre-supernova models in the ($M_4\cdot\mu_4$, $\mu_4$,) plane. The blue and magenta lines are separating likely explosions (below the line) and likely black hole formation (above the line). The parametric lines are taken from \cite{Ertl_Janka_Woosley_Sukhbold_Ugliano_2016}. The blue line corresponds to the calibration using the s19.8 model (Eq.~\eqref{eq:Ertlcriterion1}) and the magenta line to the w20 model (Eq.~\eqref{eq:Ertlcriterion2}).}
    \label{fig:mu4M4}%
\end{figure}

Following \cite{Ertl_Janka_Woosley_Sukhbold_Ugliano_2016}, we apply the two parameter criterion separating unsuccessfully explosions from explosions as in Figure 8 of that paper. We choose to show two separation lines, even though the difference is marginal. First we use the line for the calibration model s19.8, see \cite{Ertl_Janka_Woosley_Sukhbold_Ugliano_2016} for further details,
\begin{equation}
\label{eq:Ertlcriterion1}
    \mu_4 = 0.274 \mu_4 M_4 +0.047,
\end{equation}
shown in blue in Fig.~\ref{fig:mu4M4}, this should be used to study the solar metallicity $15M_{\odot}$ models. Second, we use the line calibrated on the w20 model
\begin{equation}
\label{eq:Ertlcriterion2}
    \mu_4 = 0.284 \mu_4 M_4 +0.0393,
\end{equation}
shown in magenta and used to analyse our lower metallicity models $20M_{\odot}$ and  $25M_{\odot}$.
With symbols, we plot in Fig.~\ref{fig:mu4M4} the location of the models of Table.~\ref{table:pre_sn_compar}. For the $25M_{\odot}$ cases the prediction for GENEC and MESA models are failed explosions, although, the MESA models all lie very close to the separation limit. For the $15M_{\odot}$ case, the GENEC model will most likely produce an explosion, whereas for the outcome of the MESA model remains more uncertain. As with other previously discussed aspects we see that small details and differences between the two codes could lead to an impact on the explosion success of the progenitors.

Let us nonetheless assume that all models form a neutron star. We can estimate the hypothetical gravitation mass of such an object following the methodology presented in \cite{Griffiths_Eggenberger_Meynet_Moyano_Aloy_2022} adapted from \cite{Hirschi_2005}. We further assume that the resulting neutron star has a radius $R=12\rm km$ and a baryonic mass of equal to $M_4$ 
\citep[following e.g.][]{Ertl_Janka_Woosley_Sukhbold_Ugliano_2016,Sukhbold_2018ApJ...860...93}. Due to the losses of neutrinos in the core, the baryonic mass will reduce until reaching the gravitational mass of the hypothetical neutron star.  We can solve for the gravitational mass of the compact object, $M_{\mathrm{4,g}}$, by resolving the following equation
\begin{align}
    \label{eq:M4g}
    0.1 G M^2_{\mathrm{4,g}} + ( Rc^2 +0.5 GM_{\mathrm{4}})M_{\mathrm{4,g}} -Rc^2M_{\mathrm{4}} = 0.
\end{align}
The results of this computation can be found in Table.~\ref{table:pre_sn_compar}. 
The gravitational masses of the hypothetical neutron stars formed by GENEC models show a similar spread to that of MESA ($1.49\,M_\odot\lesssim M_{4,g}\lesssim 1.65\,M_\odot$) except for a lighter neutron star mass in the 25$M_{\odot}$ case. 
At this level, the changes in the core masses originate from the evolutionary differences between the models promote a different pattern/history of convection in the shells surrounding the inner core, specially during the last year of evolution (Fig.~\ref{fig:Kippenhan}). The mentioned differences are mostly attributed to the different treatment of convection in MESA and GENEC. Also the exact rates used by the nuclear network impact the nuclear energy being deposited in various shells which will change the behaviour of convective shells. We see from Fig.~\ref{fig:Kippenhan} that from carbon burning onwards the central mass fractions of the most important species over time do not always agree between GENEC and MESA, suggesting different burning patterns in the core and this will also apply to subsequent shell burning above the core in the final phases. The culmination of these small differences during the advanced phase evolution eventually will lead to the discrepancies we see in the inner stellar structure at the pre-SN link.

\section{Discussion}
\label{sec:discussion}
In this work, we have adapted GENEC to evolve to the pre-supernova point including a reduced nuclear network similar to the base MESA approx21\_plus\_co56 nuclear network. The evolution of the central density and temperature for  GENEC and MESA models share almost identical slopes at the approach to collapse and in general the evolution of models shown by Fig.~\ref{fig:rhoT_all} is quite similar between codes. 
Looking to the central core value of the electron fraction; however, there is a significant spread of the final values found between codes. The approximate network used is very dependent on the rates chosen to model the electron capture chain, and the method used to tune these rates is generally heuristic. As has been pointed out (e.g. \cite{2024_renzo}) the exact value of the profile of the electron fraction in the core is extremely important for follow-up supernovae studies and small variations here may lead to large variations in the post-collapse evolution. The approximate electron capture chain is  however enough to produce values at pre-collapse within the range of values found by \cite{Farmer_2016} in their network and mass resolution study. Depending on the purpose of the models, the network used here may suffice. For instance, to explore further with three-dimensional studies the topology of the magnetic field and the vigour of convection in the inner regions of the star pre-collapse, this reduced network is, perhaps, the only practical one. Naturally, a detailed nucleosynthesis forecast requires the use of larger networks to also include the decrease in the electron fraction during neon and oxygen burning. This reduces the starting value of $Y_e$ at the beginning of silicon core burning, which can change from roughly 0.49 to $\approx$ 0.48 before even beginning the electron capture chain that we use in GeValNet25. One further point regarding networks is the use of approximate energy production. In this work, as explained in Sect.~\ref{sec:Network}, for silicon burning the energy is not computed from the full network but rather by assuming QSE and thus only accounting for the energy produced by the reaction $\rm ^{44}Ti(\alpha,\gamma)\rm ^{48}Cr$ (for isotopes heavier than $\rm ^{28}Si$). This approximation is introduced to improve the stability during the silicon core burning phase.  This is of course a practical approximation to employ, but using a fully coupled network that includes the full energy calculation will undoubtedly impact the progenitor model structure by changing the burning power in convective regions. Nonetheless, small reaction networks for the calculation of the energy production rate are often used to drive the evolution  as they are stable and efficient, \citep[see e.g.][]{Weaver_Zimmerman_Woosley_1978,Heger_Woosley_2010,Rauscher_2002,Chieffi_Limongi_2020ApJ...890...43}.

We have mentioned previously the variations in the treatment in convection and overshooting varies between the two codes which can be a contributing factor to the different final results. Stellar modelling is sensitive to small variations in the size of a convective shell and it has been pointed out that understanding the small differences between codes is of utmost importance to improve stellar modelling in general \citep{Chieffi_Limongi_Straniero_1998}. With this latest version of GENEC we aim to reduce as much as possible the differences with other stellar evolution codes  given identical or very similar input physics. The EoS, opacities and network all aim to incorporate the same effects as MESA to allow for a close comparison. However, differences between codes can also be numerical and not physical in origin, for example the manner of the coupling of the network to the structure equations or the manner in which one implements mixing.

\begin{figure}[ht]
    \centering
    \includegraphics[width=0.48\textwidth]{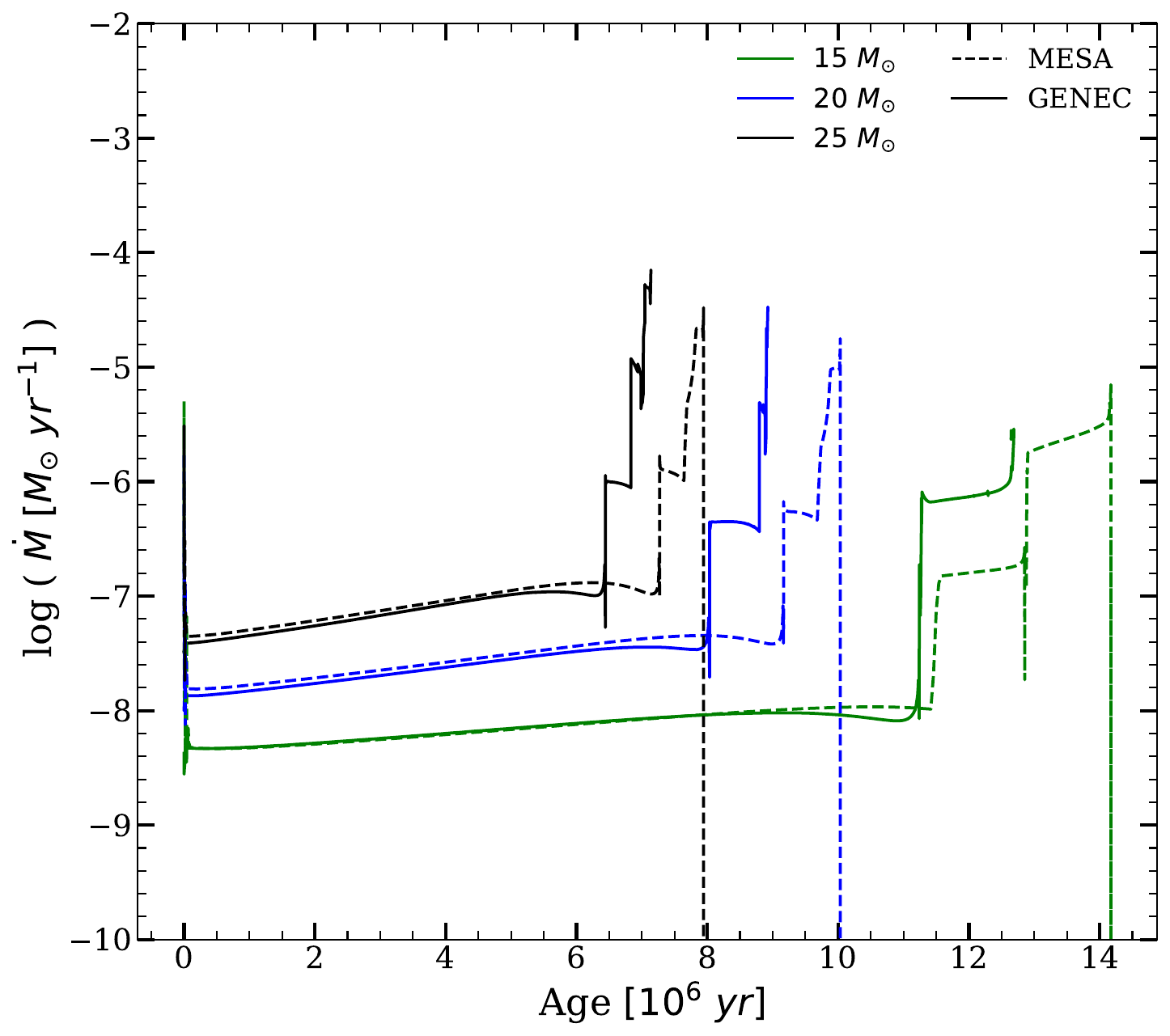}
    \caption{Mass loss rate as a function of age for all models. For each model, the main sequence lasts until the first jump in mass loss rate. The end of each curve is the age of the star at collapse.}
    \label{fig:Mdot}
\end{figure}

In Table.~\ref{table:pre_sn_compar} we see that the final masses and radii are different between codes. There exist differences in the implementation of mass loss between GENEC and MESA, particularly when dealing with the treatment of super Eddington luminosity cases, which happens for the $25M_{\odot}$ model. However, we show in Fig.~\ref{fig:Mdot} the mass loss history and during the main sequence and red giant phase the amplitude of the mass loss between codes are very similar. The difference in mass loss history rather comes from the duration of each of the different phases, where with MESA the stars tend to live longer in the main sequence and helium burning, thus leaving behind smaller total masses\footnote{The exception being the $25M_{\odot}$ model that due to the treatment of mass loss being amplified in the super Eddington luminosity case loses mass faster than MESA reaching almost $10^{-4}M_{\odot} \ \rm yr^{-1}$. }. 
The impact of mass loss is  important for surface properties, such as surface abundances, or in the case of rotating stars, the surface velocities. However, for this study, where we are mainly focused on core properties and the final shell structure deep in the star, the mass loss is of much less significance. To support this argument, we refer to \cite{Farmer_2016}, who compared models with and without mass loss, finding only a small impact on core properties.

We would finally like to highlight the variation in the Si/O shell structure between masses and codes
%models 
(see Sect.~\ref{sec:precol}). The precise nature of the region encompassing the Si-shell and the beginning of the oxygen shell can play an important role in studying supernova explosions. As pointed out by \cite{Yoshida_Takiwaki_Kotake_Takahashi_Nakamura_Umeda_2019}, the amount of convective turbulence that a multi-D model evolved for a few minutes before collapse may produce in the silicon shell region strongly depends on the coexistence  of silicon and oxygen. At the point of pre-supernova link, all models between codes and masses have very different Si/O shell structures. Some are in the process of merging, some are already merged and others well divided still. These different cases lead to convection being active in vastly different regions between models. With more detailed 1D studies and also multidimensional follow-ups of the pre-collapse evolution, we will be able to determine the origin and the importance of the differences seen in the shell structure between codes at the pre-SN link.

One final technical point may impact the very late evolution of the core (100's of seconds before collapse). Namely, the choice of the outer radius of the innermost mass shell (a parameter for GENEC). If it is too large during silicon core burning, the convective cores can be artificially bloated by the lack of mass resolution close the stellar centre. We noted a $\sim 0.4\,M_\odot$ difference in the iron core mass for the $25 M_{\odot}$ GENEC model when changing the size of the innermost cell from 100\,km to 10\,km. This highlights the importance of a follow-up mass resolution study for GENEC models, in the same vein as \cite{Farmer_2016}.

\section{Conclusions}
\label{sec:conclusions}

We have improved the GENEC code to operate in the regimes of high-density and temperature characteristic of the late phases of massive stellar evolution. The improvements have included various fronts. First, we have implemented the Timmes EoS, which is thermodynamically consistent and takes into account the positron contribution to the pressure. Second, we have also extended the opacity computation to be consistent with the latest developments in the literature. Finally, we have added some elementary electron capture into the nuclear reaction network to provide a basic tracking of the neutronisation of the iron core. Our results are in good general agreement with the equivalent models run using the MESA stellar evolution code, and also with existing models in the literature employing other stellar evolution codes. 

The reduction of the electron fraction in the core of the GENEC models using the GeValNet25 reaction network  is consistent with the uncertainty ranges of other studies employing various network sizes \citep[e.g.][]{Woosley_Heger_Weaver_2002,Rauscher_2002,Farmer_2016}. This reduction inevitably will lead to core collapse for the GENEC and MESA models alike, as shown by the core comparison with the Chandrasekhar mass. Thus, these models are usable to follow up on core collapse and supernova post-collapse simulations. 

Whilst the GENEC models are within general agreement with state-of-the-art evolution codes, there is some variability in the shellular structure between codes, which affects the explodability criterion and entropy profiles, among other properties. These variations may stem from subtle differences in the treatment of convection and the different nuclear reaction networks, causing gradual changes throughout the evolution.

We intend to extend upon this small grid of models to have a larger coverage of the parameter space, but we also plan to include the effects of rotation and of magnetic instabilities, both of which can play crucial roles in stellar evolution and the pre-supernova model structure \citep[e.g][]{Meynet2004, 2015_wheeler, Griffiths_Eggenberger_Meynet_Moyano_Aloy_2022, Eggenberger2022}. Additionally, the rudimentary network GeValNet25 was constructed to provide a fast and approximate network that included the reduction of the electron fraction in the core. Development of an extension of this network is ongoing to give a better estimation of the nucleosynthesis of important isotopes for observational purposes and also to include more sophisticated neutrino losses during silicon core burning.

\begin{acknowledgements}
AG, MR, MAA, and MO acknowledge support from grant PID2021-127495NB-I00, funded by MCIN/AEI/10.13039/501100011033 and by the European Union “NextGenerationEU" as well as “ESF Investing in your future”. Additionally, they acknowledge support from the Astrophysics and High Energy Physics programme of the Generalitat Valenciana ASFAE/2022/026 funded by MCIN and the European Union NextGenerationEU (PRTR-C17.I1) as well as support from the Prometeo excellence programme grant CIPROM/2022/13 funded by the Generalitat Valenciana. M.O. acknowledges the support of the Spanish Ramon y Cajal programme (2018-024938-I). MR acknowledges support from the Spanish Juan de la Cierva programme (FJC2021-046688-I). GM and SE have received funding from the European Research Council (ERC) under the European Union's Horizon 2020 research and innovation programme (grant agreement No 833925, project STAREX). RH acknowledges support from the World Premier International Research Centre Initiative (WPI Initiative), MEXT, Japan, the IReNA AccelNet Network of Networks (National Science Foundation, Grant No. OISE-1927130), the European Union’s Horizon 2020 research and innovation programme (ChETEC-INFRA, Grant No. 101008324) and the Wolfson Foundation.

\end{acknowledgements}
%\newpage
\bibliographystyle{aa}
\bibliography{biblio.bib} 
\label{LastPage}
%\end{document}
\begin{appendix}

\section{The GENEC code}
\label{sec:GENECbasics}
The stellar evolution code GENEC solves the internal structure equations as described in the appendix~A of \cite{Meynet_Maeder1997} and where the numerical methods employed are detailed in  \cite{Kippenhahn_1967MComP...7..129}. A complete description of the code (without the improvements included in this paper) can be found in \cite{2008_Eggenberger}. The discretisation of the equations follows the prescription of \cite{Sugimoto_1970}, but does not include any acceleration term (i.e. they assume hydrostatic equilibrium). The nuclear reaction network during the main sequence is essentially the same as employed in \cite{Meynet_Maeder2003}, and up to carbon burning it follows closely the implementation done in the PhD thesis \cite{2004_Thesis_Hirschi}, with the updates in reaction rates included in \cite{Ekstrom_Georgy_Eggenberger_Meynet_Mowlavi_Wyttenbach_Granada_Decressin_Hirschi_Frischknecht_et_al._2012}. Modifications in the late phases of evolution are included here (Sect.~\ref{sec:Network}). The treatment of convection and diffusion can be found, for example in \cite{Ekstrom_Georgy_Eggenberger_Meynet_Mowlavi_Wyttenbach_Granada_Decressin_Hirschi_Frischknecht_et_al._2012} and references therein. Rotation is implemented following the basic prescription of \cite{Meynet_Maeder1997}, where the structure equations are averaged out on isobars and a Roche-approximation \citep[see, e.g.][~\S~2.1]{Maeder_2009} is employed. GENEC includes the action of meridional circulation currents, employing an advective treatment \citep[instead of diffusive; e.g.][]{Maeder_Zahn_1998}.  

Next, we present briefly some further details about the numerical efficiency of the GENEC models computed in this paper. In GENEC the typical number of mass shells during evolution is $\approx$ 1500, which correspond to an average mass resolution of $\delta m = 5 \times 10^{-3} M_{\odot}$. This is to be compared with the MESA models which use an average of 14000 mass shells corresponding to an average mass resolution of $\delta m = 5 \times 10^{-4} M_{\odot}$. We note that the number of mass shells in GENEC adapts dynamically during the evolution, adding shells where higher resolution is needed to capture strong gradients, or removing them where such resolution is no longer needed. Layers may be added or removed depending on whether various criteria are met. For example, if the variation in pressure, radius or luminosity between two cells is very small then these cells are fused, but, on the contrary, if the relative variations of any of these quantities is too large, cells are added.  

The typical number of time-steps used by a MESA model is on the order of 20000, whereas for a GENEC model this can be up to 6-7 times more. Noteworthy in GENEC simulations, more than 60\% of the steps correspond to the evolution posterior to oxygen core exhaustion, highlighting the numerical difficulty of this late phase of evolution. The small time steps needed for the evolution in that phase are produced by the stiffness in the nuclear reaction network. We point out that in the solution strategy of GENEC, the temperature is assumed constant during a burning sub-step. After it, temperature (and the rest of the variables in the Henyey iteration) are updated. Further iterations of alternating sub-steps of burning and of the Henyey solver are performed until convergence is found and a time step is completed. During the late phases of evolution, this strategy limits significantly the time step size for stability, as well as for accuracy. We remark that the small steps taken in GENEC are also enforced to maintain the sum of mass fractions as close to one as possible, with an absolute tolerance $1-\sum_i X_i < 10^{-5}$. That is enforced without renormalising mass fractions after any burning step. In contrast, MESA, solves the nuclear network along with the temperature, allowing for renormalisation of the isotopic abundances
(potentially numerically modifying these abundances during the normalisation), providing longer time steps over which the solution is stable.
%(but, at the cost of not preserving the elementary masses). 
Future versions of GENEC may consider this solution strategy to improve the numerical performance of the code.

\section{The Dichte EoS}
\label{sec:Dichte_EOS}

The Dichte Equation of State, which up until now was the only EoS implemented in GENEC, considers stellar matter as a mixture of perfect ion gas, free electrons and radiation. It further takes into account partial ionisation of the atoms and partial degeneracy of the electrons. In this section we explain in detail the functioning of this EoS and notably the manner in which it takes into account partial ionisation in the stellar envelope. We recall that in the latest models presented in this paper the T-EoS is only used where the condition of Eq.~\ref{eqn:EOS_cond} holds, otherwise we resort to the D-EoS.

The pressure in the stellar interior is composed of two parts, gas pressure ($P_{\rm g}$) and radiative pressure ($P_{\rm r}$),
\begin{equation}
    P = P_{\rm g} + P_{\rm r} = P_{\rm g}+ \frac{1}{3}aT^4.
\end{equation}
The gas pressure itself is the sum of the ion contribution and the electron/positron contribution. The D-EoS does not take into account the positron contribution, which is negligible in the regime where we apply it currently, thus the gas pressure is written as, 
\begin{equation}
    P_{\rm gas} = P_{\rm i} + P_{\rm e},
\end{equation}
where the pressure of the ions is,
\begin{equation}
\label{eqn:P_ions}
    P_{\rm i} = \frac{\mathcal{R}}{\mu_0}\rho T,
\end{equation}
with $\mathcal{R}$ the gas constant per unit mass, and $\mu_0^{-1} =\sum_i X_i/A_i$ the mean molecular weight of the non-ionised matter. In the case where electrons are non-degenerate the electron pressure takes a similar form to Eq.~\ref{eqn:P_ions},
\begin{equation}
    P_{\rm e} = \frac{\mathcal{R}}{\mu_0}E\rho T,
\end{equation}
where $E$ is the number of electrons, per atom, released by ionisation. In the case of partial ionisation, we can find the value of $E$ using the Saha equations. This regime is only reached in the stellar envelopes and is applied only to a selection of the lighter elements common in this region, (i.e H, He, C, O, Ne, and Mg), the physics of this module was developed in \cite{1992_Schaller}. Here, to illustrate the method and for the purpose of simplicity, we explicitly write the equations for the case where only hydrogen and helium are considered. The fraction of the number of atoms of hydrogen and helium present in the mixture are,
\begin{equation}
    \nu_1 = \nu_{H} = \mu_0 X \hspace{1.5cm} \nu_2 = \nu_3 = \nu_{He} = \mu_0 \frac{Y}{4}.
\end{equation}
Where $X$ and $Y$ are the mass fractions of hydrogen and helium, respectively. We denote by $x_1=x_H$, $x_2=x_{He}$, $x_3=x_{He^+}$ the degrees of ionisation of hydrogen and helium (once and twice ionised), respectively. Thus, we can express $E$ as,
\begin{equation}
    E = \nu_1 x_1 + \nu_2x_2 + \nu_3 x_3.
\end{equation}
The ionisation degrees, $x_i (i=1,2,3)$ verify the Saha equations,
\begin{align}
    \frac{x_1}{1-x_1}\frac{E}{1+E} & = K_1,     \label{eq:Saha1}\\
    \frac{x_2}{1-x_2-x_3}\frac{E}{1+E} & = K_2, \label{eq:Saha2}\\
    \frac{x_3}{x_2}\frac{E}{1+E} & = K_3,       \label{eq:Saha3}\\
\end{align}
with,
\begin{equation}
    K_i = \frac{\omega_i}{P_{\rm gas}} \frac{(2\pi m_u)^{3/2}(kT)^{5/2}}{h^3}e^{-\chi_i/kT},\: i=1,\ldots,3.
\end{equation}
Here we have introduced $\omega_i$, the ratios between the partition functions of two successive ionisation states and $\chi_i$ the ionisation potential for each atom or ion. In the form written above, the Saha equations do not take into consideration the different excited states for each species and neglects the effects of electrostatic interactions. With these approximations, the partition functions are replaced by the statistical weights of the ground states of the atoms or ions, so we have that,
\begin{equation}
    \omega_i =  \begin{cases}
      1 & \text{for hydrogen, }\\
      4 & \text{for helium, }\\
      1 & \text{for ionised helium, }
    \end{cases}
\end{equation}
and,
\begin{equation}
    \chi_i =  \begin{cases}
      13.598 \ \rm eV & \text{for hydrogen, }\\
      24.587  \ \rm eV & \text{for helium, }\\
      54.416 \ \rm eV & \text{for ionised helium. }
    \end{cases}
\end{equation}

With this set of equations one may solve iteratively the Saha equations \eqref{eq:Saha1}-\eqref{eq:Saha3} and determine the value of $E$, and thus the electron pressure for partial ionisation. We note that we consider complete ionisation for hydrogen and helium, ($x_1=1$, $x_2=0$ and $x_3=1$ ) when the ionisation pressure is sufficiently large so that the Saha equations are no longer valid. In practice, we assume complete ionisation if \citep[][\S\,14.6]{Kippenhahn_Weigert_Weiss_2013sse..book} 
\begin{equation}
 \rho>2.66\times 10^{-3}\mu_0\,\text{g\,cm}^{-3}  , 
 \label{eq:ionizationbypressure}
\end{equation}
which translates into the following $T-P$ relation,
\begin{equation}
    \log P - \log T + \log\beta - \log(1+E) > 5.3447, 
%    \log T - 0.123 \log P > 4.09.
\label{eq:ionizationbypressure2}
\end{equation}
where $\beta=P_{\rm g}/P$.
%
%\textcolor{red}{Have no real reference for this equation.}
In this case the electron pressure is computed by taking all species as totally ionised not just hydrogen and helium and summing their contributions to compute $E$ and $P_{\rm e}$. 
This is accurate until we reach the regime of partial degeneracy where then the electron pressure is treated differently, that is when
\begin{equation}
    \log P - \frac{5}{2} \log T + 1.6 \ge 0,
\end{equation}
which roughly corresponds to an electron degeneracy parameter $\psi=\lesssim -4$ \citep{Hofmeister_1964ZA.....59..215}.
Under the conditions of full ionisation, the electron pressure is computed using suitable approximations of the Fermi-Dirac integrals, including the  both the non-relativistic and relativistic regimes. Given an input pressure and temperature, $(P,T)$, the D-EoS assumes that,
\begin{equation}
\label{eq:P}
    P = P_\textrm{g} (T, \psi ) + P_\textrm{r}(T) = P_\textrm{g} (T, \psi ) + \frac{1}{3}aT^4.
\end{equation}
As in the regime of partial ionisation, the gas pressure results from the contribution of the ions and of the free electrons,
\begin{equation}
\label{Gas pressure}
    P_\textrm{g} (T,\psi) = \frac{\mathcal{R}T}{\mu_{0}} \rho (T,\psi) + P_\textrm{e} (T, \psi).
\end{equation}
Expressions \eqref{eq:P} and \eqref{Gas pressure} provide implicitly the degeneracy parameter $\psi$ for known $(P,T)$. Thereby, one employs an iterative technique to recover $\psi$. Numerical experience recommends using as starting value for the iterative procedure $\psi^{(0)}=-3$, and proceed evaluating
$\frac{\rho}{\mu_e}$ and $P_\textrm{e}$ by using analytic approximations of the following Fermi-Dirac integrals,
\begin{equation}
 \frac{\rho}{\mu_e} = \frac{8 \pi m_{\rm p}}{h^3}\int_0^{\infty}\frac{p^2 dp}{exp(E/kT-\psi)+1},
\end{equation}
\begin{equation}
    P_e = \frac{8 \pi}{3h^3}\int_0^{\infty}\frac{p^3 \left(\partial E/\partial p\right) dp}{exp(E/kT-\psi)+1} .
\end{equation}
Depending on the value of $\psi$ these integrals are evaluated using two different approximations. If $-4 \leq \psi \leq 7$ we are in the partially degenerate regime and use the approximate formulae of  \cite{Kippenhahn_Thomas_1964ZA.....60...19}, based upon estimation of the Fermi-Dirac integrals using Gauss-Laguerre quadrature formulae (with relative errors smaller than $\sim 2\times 10^{-4}$).

For values $\psi>7$, the D-EoS adopts the asymptotic developments of \cite{Chandrasekhar_1939} for total electron degeneracy. Once this method converges we know the pressure, temperature, density and corresponding contribution of the electronic pressure. With them, one evaluates the usual thermodynamic derivatives needed for describing stellar structure.  We compute the derivatives $\alpha$ and $\delta$ defined as,

\begin{equation}
    \alpha =  \left.\frac{\partial \rm{ln} \rho}{\partial \ln P}\right|_T = \left.\frac{\partial \ln \rho}{\partial \ln \psi}\right|_T \left(\left.\frac{\partial \ln P}{\partial \ln \psi}\right|_T \right)^{-1},
\end{equation}
\begin{equation}
\label{eqn:delta_eos}
    \delta = -\left. \frac{\partial \ln \rho}{\partial \ln T}\right|_{P} = -\left. \frac{\partial \ln \rho}{\partial \ln T}\right|_{\psi} + \alpha \left. \frac{\partial \ln P}{\partial \ln T}\right|_{\psi},
\end{equation}
along with the thermodynamic quantities $C_p$ and $\nabla_{\rm ad}$, which may be written as,
\begin{equation}
    C_p = \left.\frac{\partial u}{\partial T}\right|_P + \frac{P}{\rho T}\delta = \frac{P}{\rho T}\left[ (4-\frac{3}{2}\beta)\delta + 6(1-\beta)\right]
\end{equation}
\begin{equation}
    \nabla_{\rm ad} = \frac{P}{\rho T} \frac{\delta}{C_p} = \frac{\delta}{(4-\frac{3}{2}\beta)\delta + 6(1-\beta)}.
\end{equation}

\section{Matching between the D-EoS and the T-EoS }

\label{sec:Matching_EOS}
The transition between the D-EoS and the T-EoS, is done when condition \eqref{eqn:EOS_cond} is satisfied. 
\begin{figure}[ht]
    \centering
    \includegraphics[width=0.48\textwidth]{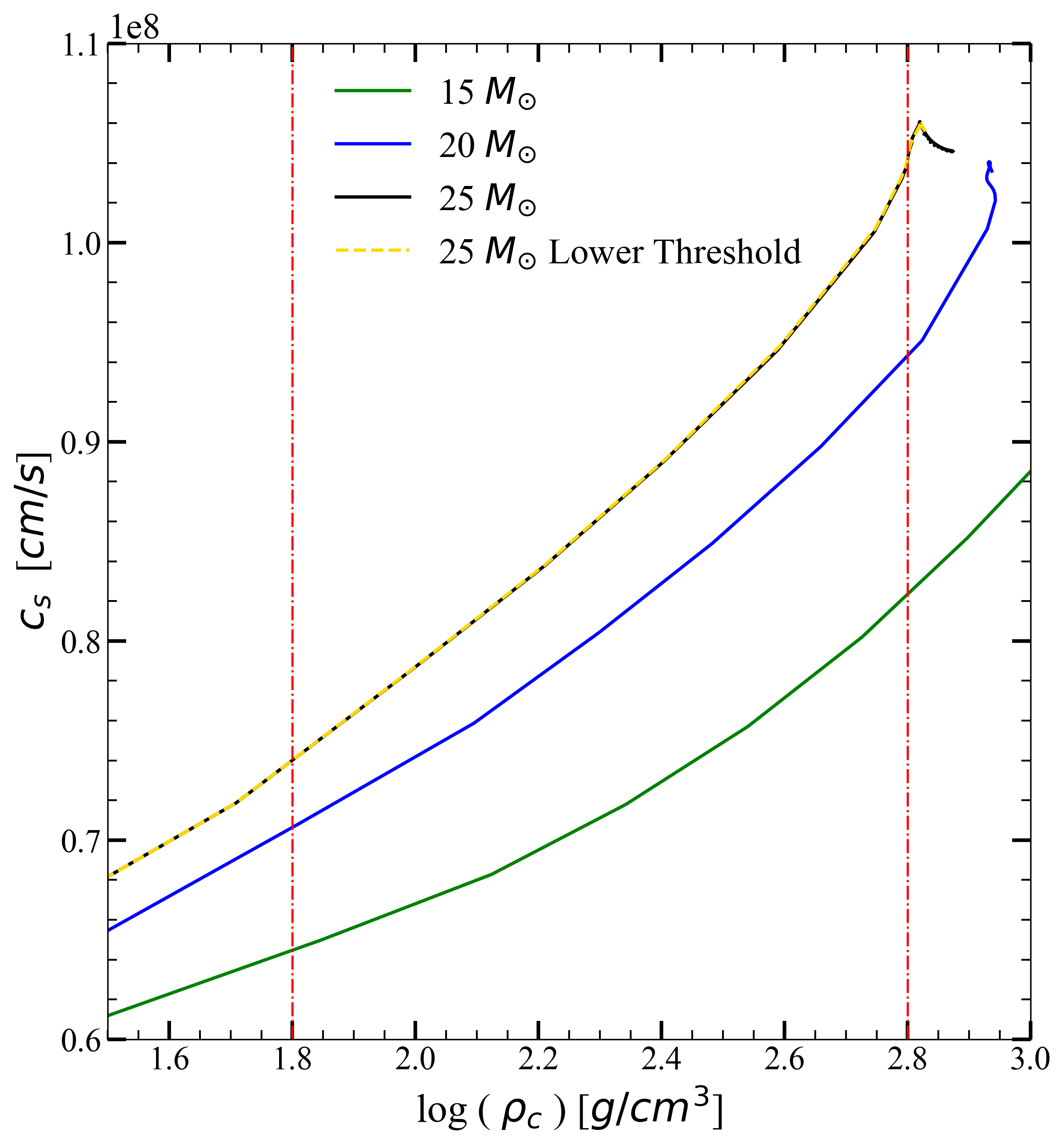}
    \caption{Evolution of the central sound speed, $c_s = \sqrt{\partial_\rho P}$, as a function of central density. The matching point between the two EoSs is at the central density value $ \log ( \rho_c ) = 2.8$, except for the lower threshold model which changes EoS at $\log ( \rho_c ) = 1.8$. }
    \label{fig:cs_evol}
\end{figure}
To demonstrate that the matching point does not have spurious effects on the models, we plot the central value of the sound speed and of $\delta$ (Eq.~\ref{eqn:delta_eos}) as a function of central density. In Figs.~\ref{fig:cs_evol} and \ref{fig:delta_evol},  we can see the smooth transition of these thermodynamic derivatives through the density threshold located at $\log(\rho_{\rm matching}) = 2.8$. We note that for the $25 M_{\odot}$ model (black line) there is a sharpness close to the threshold, this is not due to the EoS change but rather the evolutionary stage it has reached at that point. To demonstrate this, we show how the evolution of the same $25M_{\odot}$ model but using a lower density threshold ($\log(\rho_{\rm matching}) = 1.8$; dashed yellow line) overlaps the evolution of the former model with the standard transition threshold in the case of the sound speed (Fig.~\ref{fig:cs_evol}). The evolution of $\delta$ (Fig.~\ref{fig:delta_evol}) shows a tiny difference after crossing the lower density threshold, which tends to disappear once the central density rises by one order of magnitude. Thus, the matching of the two equations of state here is robust and does not produce numerical artefacts that could be troublesome for the computation.

\begin{figure}[ht]
    \centering
    \includegraphics[width=0.48\textwidth]{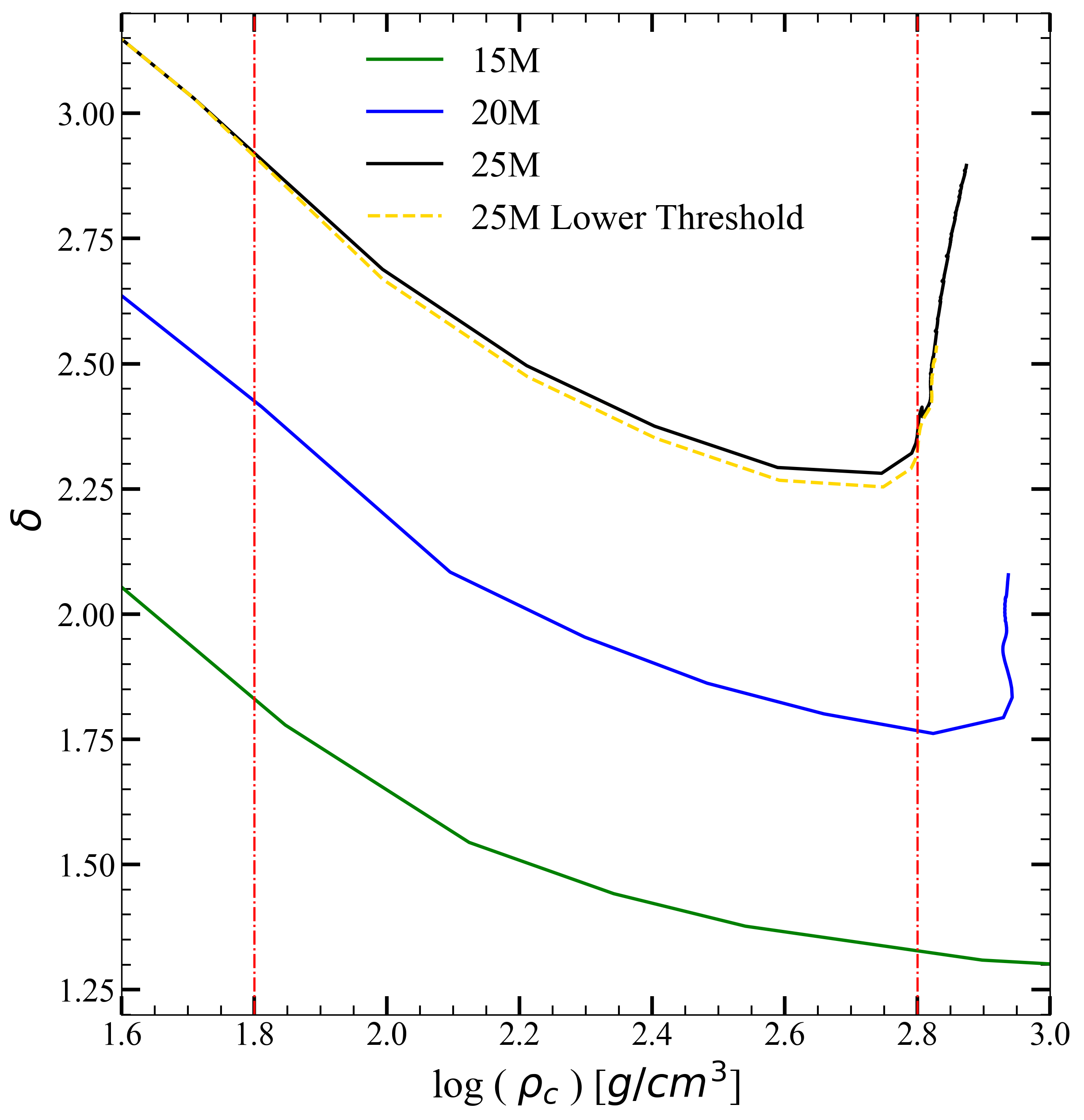}
    \caption{Evolution of the thermodynamic derivative $\delta$, defined in Eq.~\ref{eqn:delta_eos}, computed at the centre of the star as a function of central density. The matching point between the two EoSs is at the central density value $ \log ( \rho_c ) = 2.8$, except for the lower threshold model, which changes EoS at $ \log ( \rho_c ) = 1.8$.}
    \label{fig:delta_evol}
\end{figure}

\end{appendix}

\end{document}